\documentclass[fleqn,usenatbib]{mnras}

\usepackage[T1]{fontenc}

\DeclareRobustCommand{\VAN}[3]{#2}
\let\VANthebibliography\thebibliography
\def\thebibliography{\DeclareRobustCommand{\VAN}[3]{##3}\VANthebibliography}

\usepackage{graphicx}	
\usepackage{amsmath}	
\usepackage{amssymb}	
\usepackage{newtxtext,newtxmath}

\def\msun{\hbox{M$_\odot$}}
\def\vsini{\hbox{$V\sin i$}}

\def\nstars{2\,184}

\title[Stellar rotation in NGC~1850]{The effects of stellar rotation along the main sequence of the 100~Myr old massive cluster NGC~1850\thanks{Based on observations collected at the European Organisation for Astronomical Research in the Southern Hemisphere under ESO programmes 0102.D-0268 and 106.216T.}}

\author[S. Kamann et al.]{
S. Kamann,$^{1}$\thanks{E-mail: s.kamann@ljmu.ac.uk}
S. Saracino,$^{1}$
N. Bastian,$^{2,3}$
S. Gossage,$^{4}$
C. Usher,$^{5}$
D. Baade,$^{6}$
I. Cabrera-Ziri,$^{7}$\newauthor
S.~E. de Mink,$^{8,9}$
S. Ekstrom,$^{10}$
C. Georgy,$^{10}$
M. Hilker,$^{6}$
S.~S. Larsen,$^{11}$
D. Mackey,$^{12}$
F. Niederhofer,$^{13}$\newauthor
I. Platais,$^{14}$
D. Yong$^{12}$
\\
$^{1}$Astrophysics Research Institute, Liverpool John Moores University, IC2 Liverpool Science Park, 146 Brownlow Hill, Liverpool L3 5RF, UK\\
$^{2}$Donostia International Physics Center (DIPC), Paseo Manuel de Lardizabal, 4, 20018, Donostia-San Sebasti\'an, Guipuzkoa, Spain\\
$^{3}$IKERBASQUE, Basque Foundation for Science, 48013, Bilbao, Spain \\
$^{4}$Center for Interdisciplinary Exploration and Research in Astrophysics (CIERA), Northwestern University, 2145 Sheridan Road, Evanston, IL 60208, USA \\
$^{5}$The Oskar Klein Centre, Department of Astronomy, Stockholm University, AlbaNova, SE-10691 Stockholm, Sweden \\
$^{6}$ESO, European Southern Observatory, Karl-Schwarzschild-Str. 2, 85748 Garching bei M\"unchen, Germany \\
$^{7}$Astronomisches Rechen-Institut, Zentrum f\"ur Astronomie der Universit\"at Heidelberg, M\"onchhofstra{\ss}e 12-14, D-69120 Heidelberg, Germany\\
$^{8}$Max Planck Institute for Astrophysics, Karl-Schwarzschild-Stra{\ss}e 1, 85748 Garching, Germany \\
$^{9}$Anton Pannekoek Institute for Astronomy, University of Amsterdam, Science Park 904, 1098XH Amsterdam, The Netherlands \\
$^{10}$Department of Astronomy, University of Geneva, Chemin des Maillettes 51, 1290 Versoix, Switzerland\\
$^{11}$Department of Astrophysics/IMAPP, Radboud University, P.O. Box 9010, 6500~GL Nijmegen, The Netherlands\\
$^{12}$Research School of Astronomy and Astrophysics, Australian National University, Canberra, ACT 0200, Australia \\
$^{13}$Leibniz-Institute for Astrophysics, An der Sternwarte 16, 14482 Potsdam, Germany \\
$^{14}$Department of Physics and Astronomy, Johns Hopkins 
University, 3400 North Charles Street, Baltimore, MD 21218, USA \\
}

\date{Accepted XXX. Received YYY; in original form ZZZ}

\pubyear{2022}

\begin{document}
\label{firstpage}
\pagerange{\pageref{firstpage}--\pageref{lastpage}}
\maketitle

\begin{abstract}
Young star clusters enable us to study the effects of stellar rotation on an ensemble of stars of the same age and across a wide range in stellar mass and are therefore ideal targets for understanding the consequences of rotation on stellar evolution. We combine MUSE spectroscopy with HST photometry to measure the projected rotational velocities (\vsini{}) of 2\,184 stars along the split main sequence and on the main sequence turn-off (MSTO) of the 100 Myr-old massive ($10^5\,{\rm M_{\odot}}$) star cluster NGC~1850 in the Large Magellanic Cloud. At fixed magnitude, we observe a clear correlation between \vsini{} and colour, in the sense that fast rotators appear redder. The average \vsini{} values for stars on the blue and red branches of the split main sequence are $\sim100\,{\rm km\,s^{-1}}$ and $\sim200\,{\rm km\,s^{-1}}$, respectively. The values correspond to about $25-30\%$ and $50-60\%$ of the critical rotation velocity and imply that rotation rates comparable to those observed in field stars of similar masses can explain the split main sequence. Our spectroscopic sample contains a rich population of $\sim$200 fast rotating Be stars. The presence of shell features suggests that 23\% of them are observed through their decretion disks, corresponding to a disk opening angle of 15~degrees. These shell stars can significantly alter the shape of the MSTO, hence care should be taken when interpreting this photometric feature. Overall, our findings impact our understanding of the evolution of young massive clusters and provide new observational constraints for testing stellar evolutionary models.
\end{abstract}

\begin{keywords}
stars: rotation -- galaxies: star clusters: individual: NGC~1850 -- Hertzsprung-Russell and colour-magnitude diagrams
\end{keywords}



\section{Introduction}

Contrary to the Milky Way, the Large and Small Magellanic Cloud are known to host populations of young ($<1$~Gyr) and intermediate-age (few Gyrs) massive stellar clusters.  Due to their high masses ($\sim10^5\,{\rm M_\odot}$) and low extinction, along with their relative proximity allowing for high precision photometry, these clusters have opened new windows into a plethora of unexpected phenomena. Studies of their colour magnitude diagrams (CMDs) revealed the extended main sequence turn-off (MSTO) phenomenon \citep[e.g.,][]{2007MNRAS.379..151M}, high fractions of Be stars \citep[][]{feast72,grebel92,bastian17, milone18_be}, as well as split or dual main sequences \citep[e.g.,][]{milone_1866}.  The characteristics of these features strongly depend on cluster age \citep[e.g.,][]{niederhofer2015}, and the fact that some of them, such as the extended MSTO \citep{cordoni2018}, have also been identified in lower-mass open clusters in the Milky Way, suggests that the underlying mechanisms work irrespective of cluster masses, metallicity or birth environments.

While there are similarities to the CMDs of the ancient globular clusters, which host multiple populations, the origin of these complex features appears to be different between the old and young clusters.  In the ancient clusters, the complex CMDs are caused mainly by star-to-star chemical abundance variations \citep[see reviews by][]{BL18,gratton_review}.  Similar variations have been observed in massive Magellanic Cloud clusters with ages of $\sim2\,{\rm Gyr}$ or above \citep[e.g.,][]{saracino2020,martocchia2020,cadelano2022}, while they appear to be absent in the stars observed in even younger clusters \citep[e.g.,][]{mucciarelli14}.  Instead, stellar rotation has been suggested as the dominant cause of the unexpected features in the young clusters \citep[e.g.,][]{bastian09,dantona15}.

To date, most of the spectroscopic observational works on young clusters have been focused on the MSTO, due to the relative brightness of the stars there \citep[e.g.,][]{dupree17}.  Studies of the extended MSTOs in Magellanic Cloud and Galactic open clusters have been able to directly link the spread in the CMD with the rotation rates of the stars \citep[e.g.,][]{bastian18_n2818,marino18_m11, sun19, kamann_ngc1846}. The cause of the split main sequence is less well constrained in such clusters.  \citet{marino18} measured the projected rotation rates (\vsini{}) of 31 stars along the blue and red main sequences in the young LMC cluster NGC~1818 ($\sim40~{\rm Myr}$) and found that the average \vsini{} values of the two sequences were approximately $70\,{\rm km\,s^{-1}}$ and $200\,{\rm km\,s^{-1}}$, respectively.

The physical mechanism creating the different rotation rates is still under debate. The idea that all stars are born as fast rotators before tidal torques in binaries create the population of slow rotators \citep[e.g.,][]{dantona15} appears in conflict with observations of massive clusters in the Magellanic Clouds that have found comparable binary frequencies for fast and slowly rotating stars \citep[][]{kamann_ngc1846,kamann21}. Alternatively, \citet{bastian20} proposed that bimodal rotational velocity distributions originate from differences in the lifetimes of the stars' pre-main sequence disks while very recently, \citet{wang2022} suggested stellar mergers as the cause of the blue main sequence.

In addition to the split main sequence observed in many young massive clusters (YMCs), such clusters also contain a high fraction of classical Be stars, i.e., rapidly rotating stars with a decretion disk surrounding them \citep[e.g.][]{lee1991,rivinius2013}.  Such stars can be identified photometrically or spectroscopically through the bright H$\alpha$ emission from the disk.  The high Be fractions of up to $\sim50\%$ \citep{bastian17,milone18_be,Bodensteiner20} within YMCs with ages up to a few $100\,{\rm Myr}$ are consistent with expectations of a large rapidly rotating population of stars within the clusters as inferred through the split main sequences.  Such a large population of Be stars, all with the same age and distance, offers the opportunity to study the phenomenon in more detail, and with greater ease, than possible in the field.  For example, it allows us to study a special class of Be stars, known as shell stars, which are seen nearly edge-on and can be used to estimate the opening angle of the decretion disks.

Additionally, while there is general agreement that rapid rotation is a necessary condition for the Be star phenomenon, additional processes appear to be required for a star to become a Be star \citep[e.g.,][]{baade2020}.  There are thought to be two main channels to form Be stars, the single star evolutionary path \citep[e.g.,][]{fabregat2000} and a path through mass transfer binaries \citep[e.g.,][]{pols1991}.  The mechanism underlying the single star evolutionary path is the contraction of the stellar core during the main sequence lifetime, in combination with a radially outward transport of angular momentum to the expanding stellar envelope.  This causes the star to approach its critical rotation velocity as it approaches the MSTO \citep[e.g.,][]{ekstrom08,granada2013,hastings20}.  At some point the critical velocity approaches the rotational velocity of the star and the formation of a Keplerian decretion disk is triggered.  Alternatively, during the mass-transfer phase of a binary system, angular momentum is transferred as well, which can result in a rapidly rotating star \citep[e.g.,][]{demink2013,shao2014,klement2019}.

Observational studies have shown that binaries do play a significant role in creating Be stars.  While optical spectroscopy struggles to confirm the binary nature of Be stars owing to the low masses and small sizes of their companions \citep[e.g.,][but also see \citealt{naze2022}]{Bodensteiner20}, other methods, such as spectral energy distribution modelling of the disks \citep{klement2019}, far-UV spectroscopy \citep{wang2021}, the search for runaway Be stars \citep{boubert2018}, or near-infrared interferometry \citep{klement22} have shown that a significant fraction of Be stars lives in binaries. Furthermore, spectroscopic observations of Be stars did not reveal the enhanced surface nitrogen abundance predicted for stars stemming from the single star evolutionary path \citep[e.g.,][]{lennon05}.  For the YMCs in the Magellanic Clouds, \citet{hastings21} estimated that by adopting rather extreme assumptions regarding the initial binary population, non-canonical initial stellar mass functions, and the efficiencies of mass transfer, binaries can account for the large fractions of Be stars observed near the turn-offs of the clusters \citep[e.g.][]{bastian17,milone18_be}.  For a sample of Galatic open clusters, \citet{mcswain05} estimate that up to 73\% of the observed Be stars originated in binaries, while the rest were contributed by the single evolutionary path.  Models that invoke the single or binary star path to form Be stars make explicit predictions that can be tested with targeted observations \citep[e.g.,][]{granada2013}.

Improving our understanding of the role of rotation (and Be stars in particular) is not only of interest regarding the morphology of cluster CMDs. Decretion disks provide interesting constraints for angular momentum transport in stellar models \citep[e.g.]{rimulo18}. Furthermore, Be stars are known to occur in Be/X-ray binaries \citep{reig2011}, systems where a compact object (typically a neutron star) orbits a Be star and emits X-rays each time it passes through the disk. Such systems are of renewed relevance because they represent an intermediate step in proposed pathways for the formation of gravitational wave sources. 

With the aim to better understand the stellar populations within YMCs, we have initiated a multi-epoch observational campaign targeting NGC~1850, a $\sim100$~Myr, $\sim10^5$~\msun\ cluster in the LMC.  For this we use the Multi-Unit Spectroscopic Explorer \citep[MUSE][]{2010SPIE.7735E..08B} instrument on the VLT, which has the power to study thousands of stars, from the main sequence through the red giant branch (RGB), in a single pointing, with high enough spectral resolution to find and study binaries through their radial velocity variations as well as derive the \vsini{} values of individual stars.  Parts of this large dataset have already been used to determine the binary frequency on each of the arms of the split main sequence in order to test its origin \citep[][]{kamann21} as well as to study the peculiar system NGC~1850-BH1 \citep{saracino2022,elbadry22}. In an independent analysis of the same data, \citet{sollima22} determined a dynamical mass of the cluster of $10^{4.84}\,M_{\rm \odot}$ and found a link between oxygen abundances and line widths among MSTO stars that the authors interpreted as evidence for different stellar rotation rates. Here, we present the full data set and use it to study the rotational velocity distribution of the stars as well as the Be star population within the cluster.

This paper is organised as follows. We describe the MUSE observations and their reduction in Sect.~\ref{sec:reduction}, followed by a summary of the photometric and spectroscopic data analysis in Sect.~\ref{sec:analysis}. In Sect.~\ref{sec:stellar_parameters}, the measurement of the stellar parameters is described before the \vsini{} values are discussed in Sect.~\ref{sec:rotation}.  Sect.~\ref{sec:be_and_shell} is devoted to the Be and shell stars discovered in the data. We conclude in Sect.~\ref{sec:conclusions}.

\section{Observations and data reduction}
\label{sec:reduction}

\begin{figure*}
    \centering
    \includegraphics[width=.45\textwidth]{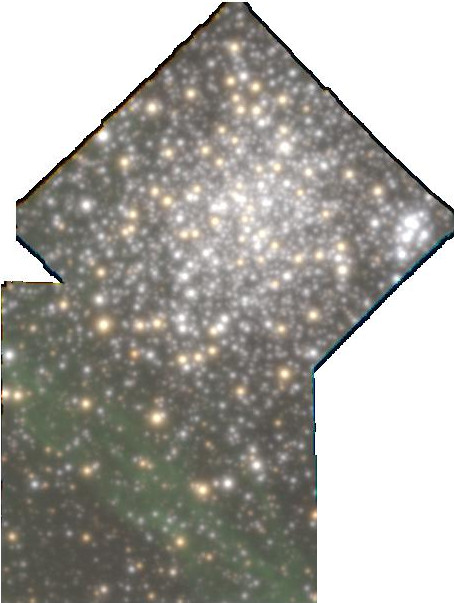}
    \includegraphics[width=.45\textwidth]{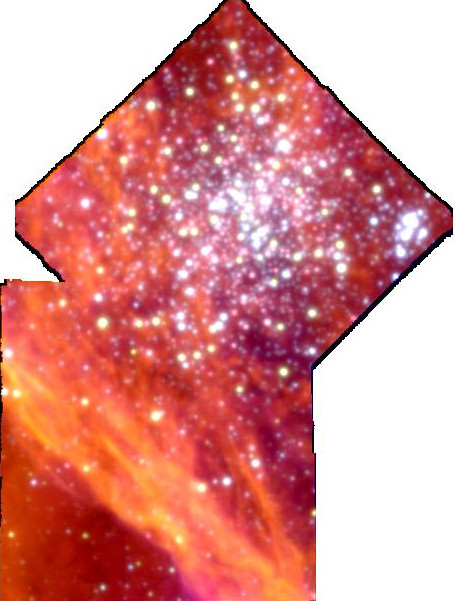}
    \caption{\textit{(Left)} SDSS-gri image created from the MUSE data of NGC~1850. \textit{(Right)} Narrow-band (\ion{O}{iii}, \ion{N}{ii}, H$\alpha$) image created from the same data to better visualise the nebular emission overlaid on the cluster. In both panels, north is up and east is left. The young cluster NGC~1850B is visible as a group of bright stars towards the western edge of the northern MUSE pointing. For this visualisation only, the two MUSE pointings have been stitched together using \textsc{Swarp} \citep{2002ASPC..281..228B}.}
    \label{fig:muse}
\end{figure*}

NGC~1850 was observed with MUSE \citep{2010SPIE.7735E..08B} at the ESO Very Large Telescope as part of programs 0102.D-0268 and 106.216T (PI: Bastian). 
The observations were carried out in the MUSE wide-field mode (WFM), which provides low- to medium-resolution spectroscopy ($R\sim1700-3500$) across a wide wavelength range ($\lambda=4\,800-9\,300\,{\text \AA}$) and over a continuous field of view of $1\arcmin\times1\arcmin$, with a spatial sampling of $0.2\arcsec$. We made use of the adaptive optics (AO) system of MUSE in order to improve the spatial resolution of the data and facilitate the extraction of single-star spectra.

The observations were carried out during 16 visits in 13 individual nights between March 2019 and February 2021. During each visit, the same two fields were observed, a central field, situated on the cluster centre and an outer field, located about $1\arcmin$ to the south-east of the cluster centre (cf. Fig.~\ref{fig:muse}). The exposure times per visit were $2\times400$~s for the central field and $3\times500$~s for the outer field. In between the individual exposures, derotator offsets of $90$~degrees and small spatial offsets of $\leq 0\farcs4$ were applied in order to homogenise the data quality across the field of view.

The data reduction was performed using versions 2.6 and 2.8.3 of the standard MUSE pipeline \citep{muse_pipeline}. All basic reduction recipes were carried out as explained in \citet{kamann_ngc1846}. However, owing to the strong nebular emission across NGC~1850, the sky subtraction procedure had to be optimized compared to our previous work. Briefly, under the \texttt{skymethod=model} setting that we adopted, the MUSE pipeline identifies the requested fraction of faintest spaxels in the field of view and combines their spectra. The result is split up into a continuum and a telluric emission-line component. While the former is subtracted globally from the data, the latter is fitted to the data of each slice individually, in order to account for variations in the line spread function (LSF) across the field of view \citep[see][for details]{muse_pipeline}. In our case, however, the continuum component always contains a contribution from the nebular emission, which would be subtracted from the data when using the standard procedure. Therefore, we manually replaced the spectrum of the continuum component with zeros. This implied that our final data cubes still contain telluric continuum emission. This, however, does not affect the extraction of the spectra described in Sect.~\ref{sec:analysis} below. The reason is that the telluric component is spatially flat across the field of view, and such a component can be readily accounted for when performing spectrum extraction via point spread function (PSF) fitting.

We created individual data cubes for each pointing and each visit. For illustration purposes, we further created one data cube combining all individual exposures from program 0102.D-0268. In Fig.~\ref{fig:muse}, we show two colour images created from this combined data cube, using either the SDSS $gri$ broadband filters or custom narrow-band filters centred in the \ion{O}{iii}, \ion{N}{ii}, and ${\rm H_{\alpha}}$ emission lines. The latter illustrate the strong extended gas emission originating from the young cluster NGC~1850B, visible in the western corner of the northern (central) pointing.

\section{Data analysis}
\label{sec:analysis}

\subsection{Photometry}
\label{sec:analysis:photometry}

The HST data were taken in 2015 as part of programs 14069 (PI: Bastian) and 14174 (PI: Goudfrooij). PSF photometry was performed on the flat-field corrected, and bias-subtracted WFC3 images in the filters F275W, F336W, F343N, F438W, F656N, and F814W. This was carried out with \textsc{DOLPHOT} \citep{Dolphin16}, a modified version of \textsc{HSTphot} \citep{Dolphin00} using Jay Anderson’s PSF library. For more details on the photometry we refer the reader to \cite{Gossage19, kamann21} and references therein. Bright stars were saturated in almost all the long exposures available, hence to recover their magnitudes we made use of the shortest exposure (7~sec) in F814W. This photometric catalog was then used as a reference for the spectra extraction (see Sec.~\ref{sec:analysis:spectroscopy} below).  A $m_{\rm F336W} - m_{\rm F438W}$ vs $m_{\rm F438W}$ CMD generated from the HST photometry is shown in Fig.~\ref{fig:cmd}.

Following our earlier work described in \citet{kamann21}, we identified two samples of red and blue main sequence stars in the magnitude interval $18.0 < m_{\rm F438W} < 20.0$, where the two sequences are most clearly separated. The reader is referred to \citet{kamann21} regarding the details of this process.  In essence, we assumed that the blue main sequence stars constitute a constant fraction of $0.2$ of all stars at a given magnitude level.  The dividing line between stars we consider to lie on the red or the blue main sequence is depicted as a solid blue line in the right panel of Fig.~\ref{fig:cmd}.  \citet{wang2022} recently reported that the ratio of red to blue main sequence stars in a sample of similarly aged massive clusters is magnitude dependent \citep[see also][]{milone18_be}.  However, visual inspection of Fig.~\ref{fig:cmd} suggests that at least for the magnitude range considered in this work, assuming a constant ratio of blue to red main sequence stars does not result in a significant number of misclassified stars.

When comparing the radial distributions of the red and blue main sequence stars, we found that the former were centrally concentrated compared to the latter. A two-sided Kolmogorov-Smirnov test of the two distributions yielded a probability of $5\times10^{-12}$ that the two were drawn from the same parent sample. This finding is at odds with the analysis of \citet{correnti17}, who did not find any differences in the concentrations of the two populations.

In order to clean our sample of red main sequence stars from photometric binaries, we imposed a second selection criterion, illustrated by the dashed blue line included in the right panel of Fig.~\ref{fig:cmd}.  This second line was constructed by shifting the dividing line between the two main sequences by $\Delta (m_{\rm F438W}-m_{\rm F336W})=0.1 + 0.025\times (m_{\rm F438W}-18)$ and follows the drop in stellar density visible to the red of the red main sequence.  Stars lying redwards of this line are considered to be in binary systems with two luminous companions.  We note that this is a purely photometric selection of binaries.  Stars showing radial velocity variations were treated as described in Sect.~\ref{sec:analysis:spectroscopy} below.

We note that photometric binaries originating from the blue main sequence would overlap in CMD space with the red main sequence.  Depending on the mechanism adopted in order to explain the split main sequence, the fraction of binaries among blue main sequence stars is expected to be lower, higher, or comparable to the fraction among the red main sequence stars.  In \citet{kamann21}, we detected similar fractions of binaries for both main sequences.  Given the relative low number of blue main sequence stars, we expect any contamination of the red main sequence from ``photometrically migrating'' blue main sequence binaries to be small.

\subsection{Isochrones}
\label{sec:analysis:isochrones}

In order to derive stellar parameters from the HST photometry, we compared the data to isochrones from the MIST database \citep{mist0,Gossage19}.  The MIST isochrones are derived from the Modules for Experiments in Stellar Astrophysics \citep[MESA][]{mesa}.  Following \citet{yang2018}, we adopted an age of $100\,{\rm Myr}$, a metallicity of $[{\rm Fe/H}]=-0.24$, a distance modulus of $18.45$, and an extinction of $A_V=0.301$.  We verified that the isochrones for this set of parameters provided a good by-eye fit to the CMD shown in Fig.~\ref{fig:cmd}.

Of particular relevance for the present work is the treatment of stellar rotation in the isochrone models, which is detailed in \citet{Gossage19}. Rotation is parameterized via the parameter $\Omega/\Omega_{\rm crit}$, specifying the fractional angular velocity of a star relative to the critical value at the zero-age main sequence (ZAMS). In the MIST isochrones, $\Omega_{\rm crit}$ is defined as the limit where the centrifugal force equals the gravity of the star.  The isochrones are available for 10 discrete steps of $\Omega/\Omega_{\rm crit}$, ranging from 0 to 0.9. Gravity darkening in the models is treated following \citet{espinosa_lara2011} and its effects are accounted for via a surface-averaged modification of the luminosity and effective temperature. In reality, gravity darkening causes a viewing-angle (i.e. inclination) dependence of the observed colours of a rotating star, which is not included in the models.

In Fig.~\ref{fig:cmd}, we compare the HST photometry to four isochrones with different rotation parameters. We note that in the magnitude range displaying the split main sequence, i.e. $18 < m_{\rm f438W} < 20$, some features of the CMD are not accurately represented by the isochrones. For example, the blue main sequence shows bluer colours than even the isochrone without rotation (i.e. $\Omega/\Omega_{\rm crit}=0$) predicts.  Recently, \citet{wang2022} argued that stellar mergers can account for this colour offset, as they lead to a rejuvenation of the merger product, resulting in bluer colours.  However, it is unclear if mergers during the first few Myr following the formation of a cluster, as advocated by \citet{wang2022}, can rejuvenate stars sufficiently to account for the observed colour shift.  Furthermore, at higher rotation rates, the isochrones predict slightly different slopes of the main sequence compared to what is observed. These deviations likely stem from remaining uncertainties in the modelling of massive stars and the treatment of rotation in those models. The treatment of rotation varies between stellar evolution codes, highlighting the need for more observational constraints suited to verify or reject assumptions made in the models.

\subsection{Spectroscopy}
\label{sec:analysis:spectroscopy}

\begin{figure*}
    \centering
    \includegraphics[width=0.95\linewidth]{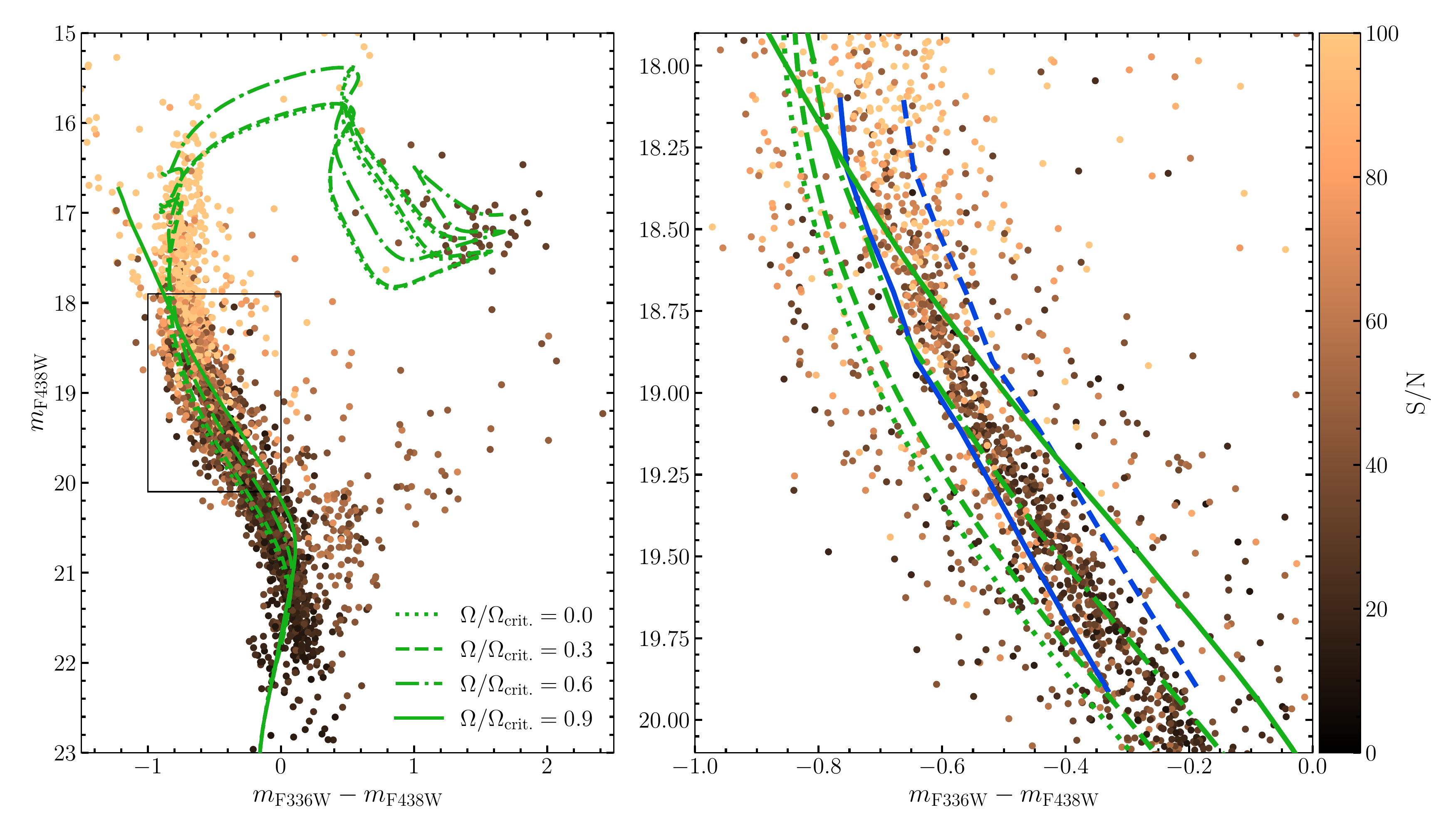}
    \caption{Colour-magnitude diagram (CMD) of NGC~1850, using HST photometry in filters F336W and F438W. The left panel shows the full CMD, and the right panel zooms into the area indicated by a black rectangle in the left panel, where the split main sequence is most obvious. In both panels, only stars with MUSE spectra are shown and colour-coded according to their average spectral signal-to-noise ratio per pixel. Green lines show the predictions from MIST isochrone models for different rotation rates, relative to the critical rotation rate $\Omega_{\rm crit}$. In the right panel, the blue solid line indicates the adopted division into red and blue main sequence stars, and the blue dashed line shows the adopted division between red main sequence stars and photometric binary stars.}
    \label{fig:cmd}
\end{figure*}

We extracted individual stellar spectra from the final MUSE data cubes using \textsc{PampelMuse} \citep{2013A&A...549A..71K}. The code works by using a reference catalogue of sources to determine a MUSE PSF model and spaxel coordinates of the resolved sources as a function of wavelength. This information is subsequently used to optimally extract the spectra of the resolved stars from the data. The reference catalogue used for NGC~1850 was the same as discussed in Sect.~\ref{sec:analysis:photometry} and the stars included in the extraction were selected based on their $m_{\rm F814W}$ magnitudes.

The extraction was performed on each individual data cube. As described in \citet{saracino2022}, the extracted spectra were thereafter analysed with \textsc{Spexxy} \citep{spexxy}, a code which determines stellar parameters via full-spectrum fitting against a library of templates. As in our previous work on NGC~1850, we used the synthetic templates from the library presented in \citet{ferre} when running \textsc{Spexxy}.

The individual radial velocities derived from the \textsc{Spexxy} fits were used in \citet{kamann21} and \citet{saracino2022} to study the binary properties of the stars in our sample.  In this work, however, our aim is to obtain a single spectrum per star in our sample at maximum signal-to-noise (S/N).  For this reason, we combined the spectra obtained for the individual visits on a star-by-star basis.  Before combining them, each spectrum was corrected for its radial velocity as measured by \textsc{Spexxy} and thereby converted to restframe.  This correction ensures that the widths of the spectral lines, which are used below to infer the stellar rotation of our sample stars, are not dominated by the orbital motions in binary systems.  We sound a note of caution that our binary detection via radial velocity variations is biased against systems composed of equally luminous companions \citep[e.g.][]{giesers2019,bodensteiner20b} with blended spectral lines. Such SB2 or double-line binaries can still result in artificially high \vsini{} measurements. We discuss their impact in Sect.~\ref{sec:rotation:individual} below.

Furthermore, we corrected the individual spectra for atmospheric absorption prior to the combining. This is possible because \textsc{Spexxy} fits the telluric absorption bands simultaneously with the stellar features, using an internal library of atmospheric spectra \citep[see][for details]{spexxy}.

In the combining process, we weighted the individual spectra by their signal-to-noise ratio (S/N, measured per wavelength bin and averaged across the entire wavelength range). For each star, spectra with a S/N lower by a factor $<0.5$ compared to the one with the highest S/N were discarded altogether, as they tended not to improve the quality of the combined spectrum. In addition, we discarded any spectra for which the results from the \textsc{Spexxy} analysis were deemed unreliable for any of the following reasons. (1) The fit was marked as unsuccessful by \textsc{Spexxy}. (2) The S/N of the input spectrum as estimated by \textsc{Spexxy} was $<5$. (3) The velocity determined by cross-correlating the spectrum with its best fitting template deviated by $>3\sigma$ from the actual \textsc{Spexxy} result.

The gaseous emission across NGC~1850 (cf. right panel of Fig.~\ref{fig:muse}) poses a major challenge for the extraction of spectra. It varies over spatial scales that are comparable to the resolution of the MUSE data ($\gtrsim0.5\arcsec$), hence it cannot be easily deblended from the stellar emission. We experimented with setting up a fine background grid (using distances down to $10$~spaxels between the individual grid points) and including the flux of each background component in the extraction process. However, residuals from the strong nebular emission lines were still visible in some of the extracted spectra. Therefore, we designed an approach to account for any contamination by nebular lines during the analysis of the spectra. For each extracted stellar spectrum, we created a mask as follows. 
We first selected a comparison sample of 100 spectra of stars with similar photometric magnitudes as the star linked to the target spectrum. Then, we calculated the median absolute deviation (MAD) of the spectral fluxes in the comparison sample and divided the result by the square root of the median spectral fluxes in the comparison sample. The latter is done in order to account for the fact that the noise in spectral absorption lines is typically lower than in the continuum. Afterwards, we removed the continuum from the result of the previous step, using a median filter of 200 wavelength bins width. On the continuum-corrected MAD spectrum, we determined the 84th percentile in a rolling window of size 200~bins and flagged all bins as contaminated where the actual values exceeded the smoothed ones by a factor of $>3$. Finally, we removed isolated masked bins by processing the mask with a minimum filter with a window size of 3 bins.  The masks created this way were included in the spectrum analysis of the combined spectra using \textsc{Spexxy}.

We show the distribution of our targets in an HST ($m_{\rm F336W} - m_{\rm F438W}$, $m_{\rm F438W}$) colour-magnitude diagram in Fig.~\ref{fig:cmd}, where they are colour-coded according to their S/N, determined using the method of \citet{stoehr08}. The low S/N ratios of the evolved stars visible to the top right of the CMD are very likely a shortcoming of the method when applied to low-resolution spectra of cool stars, as the numerous molecular bands are mistaken as noise. Given that the evolved stars only play a very minor role in the present work, we did not make an effort to correct for this effect. We also note that for some of the brightest stars no photometry in said filters was available, because the stars were saturated in the HST images. In such cases, we made use of the $m_{\rm F606W}$ and $m_{\rm F814W}$ magnitudes recovered from the extracted spectra. They were compared to one of the isochrones overplotted in Fig.~\ref{fig:cmd}, and the missing magnitudes were copied from the nearest data point in ($m_{\rm F606W} - m_{\rm F814W}$, $m_{\rm F606W}$) space.

\subsection{Cluster membership determination}
\label{sec:membership}

\subsubsection{Field stars}
\label{sec:membership:field}

In order to clean our sample from field stars, we utilised the radial velocities derived from the \textsc{Spexxy} fits. For each star, we averaged the velocity measurements obtained across the individual epochs, using inverse-variance weighting. Prior to this process, the uncertainties of the single-epoch velocities were calibrated using the method presented in Sec.~3.4 of \citet{kamann_ngc1846}. Velocities which were deemed unreliable according to the criteria outlined in Sec.~\ref{sec:analysis:spectroscopy} above were discarded before averaging the results. Following this process, our kinematic sample consists of 4\,207 stars with available radial velocity measurements.

Cluster membership probabilities were determined under the assumption that the observed sample of stars can be described by a cluster population and a field population. Regarding the cluster population, we further made the assumptions that its surface density follows a \citet{king1962} profile with the structural parameters obtained by \citet{correnti17} and that its velocity dispersion profile can be modelled by a \citet{plummer1911} profile with central dispersion $\sigma_0$ and scale radius $a_0$. For the field population, we adopted a Gaussian velocity field with mean velocity $v_{\rm back}$ and velocity dispersion $\sigma_{\rm back}$.  We note that because we did not try to clean our sample from Milky Way foreground stars, the fit parameters obtained for the field population could be biased towards low mean velocities and high dispersion values.

For each star, a membership prior was calculated based on its distance to the cluster centre and the surface density profile we adopted. The coordinates for the cluster centre were taken from \citet{milone18_be}. Then, we used \textsc{emcee} \citep{emcee}, a \textsc{Python} implementation of the affine-invariant Markov-chain Monte Carlo (MCMC) sampler presented by \citet{goodman2010}, to determine the model parameters ($v_{\rm back}$, $\sigma_{\rm back}$, $\sigma_0$, $a_0$, and the systemic cluster velocity $v_0$) in a maximum likelihood approach. We used 200 walkers in the process and the chains were propagated for 500 steps each. Discarding the first 200 steps of each chain as burn-in, we found the following set of parameters to maximise the likelihood of the model given the MUSE radial velocities, $v_0=247.1\pm0.2\,{\rm km\,s^{-1}}$, $\sigma_0=4.1\pm0.3\,{\rm km\,s^{-1}}$, $a_0=70^{+26}_{-19}\,{\rm arcsec}$, $v_{\rm back}=252.2\pm0.8\,{\rm km\,s^{-1}}$, and $\sigma_{\rm back}=20.2\pm0.8\,{\rm km\,s^{-1}}$. For each parameter, we adopted the 50th percentile of the distribution returned by the chains as best-fit parameter, while the confidence intervals were obtained from the 16th and 84th percentiles of the same chains. When comparing our results to \citet{song2021}, who performed a similar analysis on NGC~1850, we find reasonable agreement in the parameters that appear in both models (see their Table~7).

Given our set of model parameters, we are able to assign posterior membership probabilities to all stars with radial velocity measurements available, using the method described in, e.g, \citet{watkins2013}.  The distribution of posterior membership probabilities is clearly bimodal, with 11\% of the sample having probabilities $<0.1$ and 60\% having probabilities $>0.75$. Because the vast majority of the stars with $m_{\rm F336W}-m_{\rm F438W}>0.8$ and $m_{\rm F438W}<21$ (see left panel of Fig.~\ref{fig:cmd}), which constitute the red giant branch of the LMC field population, fall into the former group, we decided to consider the 3\,737 stars with membership probabilities $>0.1$ in the subsequent analyses. We note that because the mean velocity of the cluster and field populations only differ by $5\,{\rm km\,s^{-1}}$, any velocity-based separation into field and cluster stars remains somewhat uncertain.

\subsubsection{NGC~1850B}
\label{sec:membership:ngc1850b}

As mentioned above, the MUSE footprint covers the young cluster NGC~1850B, visible in the right edge of the central pointing in Fig.~\ref{fig:muse}.  In order to identify stars belonging to this cluster, we drew a circle of $10\,{\rm arcsec}$ radius around the visually estimated cluster centre ($\alpha=05^{\rm h}08^{\rm m}39.3^{\rm s}$, $\delta=-68^\circ 45^{\prime} 45\farcs5$) and considered all stars within this circle as members of NGC~1850B.  Out of the 903 stars in the HST photometry that were identified this way, MUSE spectra are available for 140 stars. In contrast to the field stars, we decided to keep the stars associated with NGC~1850B in our sample and discuss their impact on the analysis when appropriate. Given that our selection results in a mixture of stars from NGC~1850 and NGC~1850B, we did not try to perform a dedicated analysis of the stellar content of NGC~1850B.

\section{Stellar parameters}
\label{sec:stellar_parameters}

During the spectral analysis of the combined spectra with \textsc{Spexxy}, we measured the projected rotational surface velocity \vsini{} and the effective temperature $T_{\rm eff}$, and the surface gravity $\log g$ of every star. The initial values for the latter two were determined from the comparison between the HST photometry and the MIST isochrones as outlined in Sect.~\ref{sec:analysis:photometry}. In addition, for stars for which isochrone comparison suggested a value $T_{\rm eff} < 8\,000{\rm K}$, we also included the metallicity $[{\rm Fe/H}]$ as a free parameter in the analysis. As stars above this temperature threshold do not show any significant metal lines in the MUSE spectra, the metallicity was fixed to the isochrone value for such stars. Note that the inclusion of $\log g$ in the fitted parameters deviates from our analysis for NGC~1846 \citep[c.f.,][]{kamann_ngc1846}, where this parameter was fixed to the value obtained from the isochrone comparison. While the main conclusions of our work are unaffected by this choice, we found that the \vsini{} values determined with varying $\log g$ agreed better with the values obtained via individual line fits (cf. Sec.~\ref{sec:rotation:individual}).

Following the analysis of the combined spectra, we applied several quality cuts to our sample. Besides discarding results from formally unsuccessful \textsc{Spexxy} fits, we applied a S/N cut at 20, and also dropped results from spectra for which the recovered $m_{F814W}$ magnitude deviated strongly from the corresponding value in the HST catalogue. Because the latter could be a sign that the spectrum is contaminated by nearby stars (as a result of PSF mismatches or inaccuracies in the underlying HST astrometry), we discarded spectra for which the \textit{Mag Accuracy} parameter used by \textsc{PampelMuse} was $<0.5$. In combination with the cleaning for field stars as described in Sect.~\ref{sec:membership:field}, these criteria resulted in a final sample of \nstars{} stars with valid results that will be discussed in the following.

For $T_{\rm eff}$, we find a large range of values, as expected given the large range in spectral types covered by our observations. It is worth noting that we measure a median temperature offset of $1\,003~{\rm K}$ between the samples of red and blue main sequence stars determined as outlined in Sec.~\ref{sec:analysis:photometry}. For the same range in magnitudes ($18.0 < m_{\rm F438W} < 20.0$), the MIST isochrones predict temperature changes of up to $1\,500~{\rm K}$ when increasing \vsini{} from zero to almost critical. 

The median metallicity determined from the cooler stars ($T_{\rm eff} < 8\,000$, see above) is $-0.33$, with the 16th and 84th percentiles of the distribution being located at $-0.40$ and $-0.20$. Note that the scatter in our measurements does not imply an intrinsic metallicity spread in NGC~1850, but mostly reflects our measurement uncertainties. Recently, \citet{song2021} measured a metallicity of $[{\rm Fe/H}]=-0.31$ using high resolution spectroscopy, in good agreement with the value derived in this work.  In addition, our value also agrees with the metallicity obtained by \citet{sollima22} in their analysis of the MUSE data ($-0.31\pm0.01$).

\section{Stellar rotation}
\label{sec:rotation}

\begin{figure*}
    \centering
    \includegraphics[width=.99\textwidth]{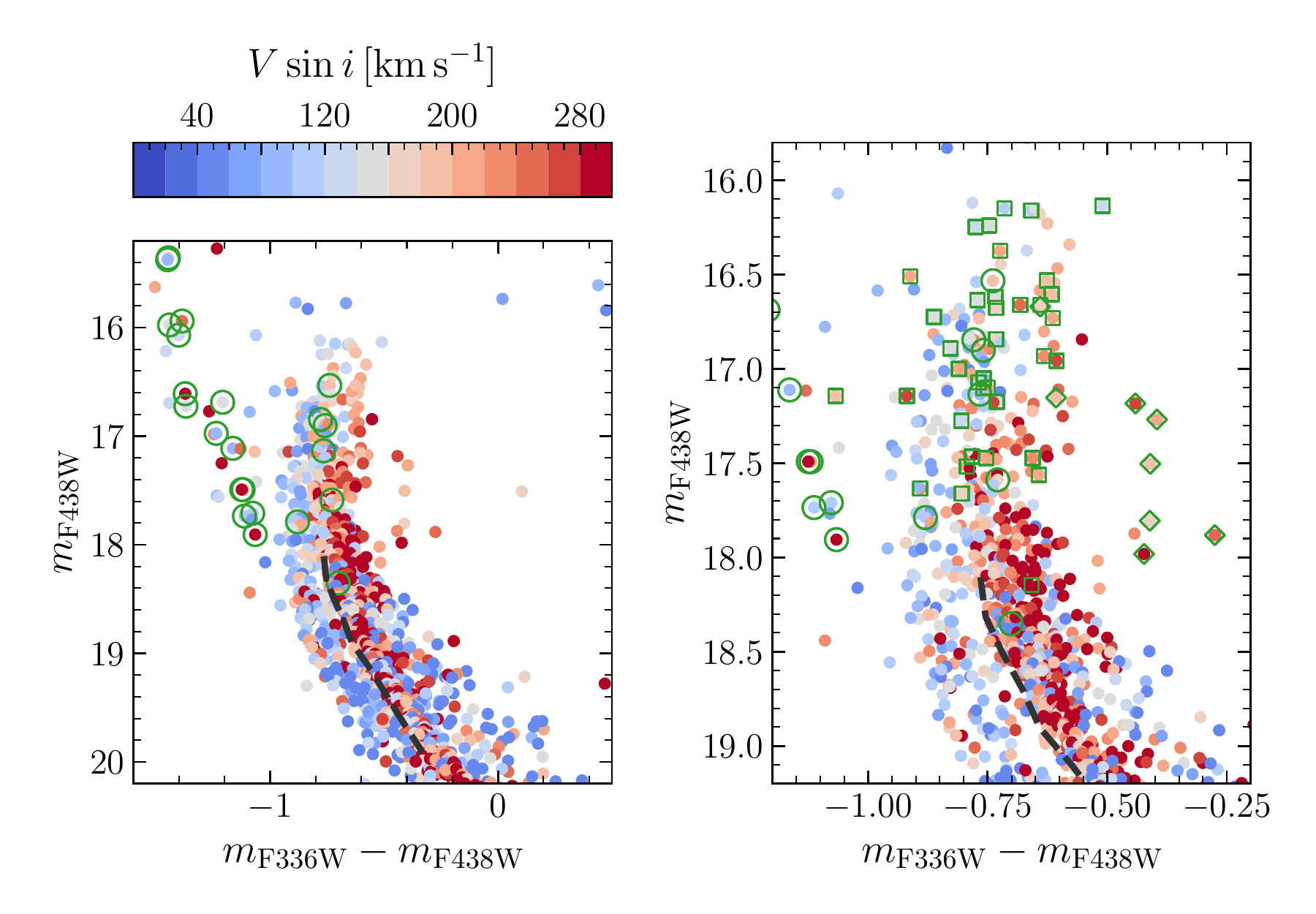}
    \caption{MUSE \vsini{} measurements for stars in NGC~1850.  On the left side, we show the HST colour-magnitude diagram of the member stars with MUSE spectra for which reliable results were obtained, colour-coded by the projected rotation velocity \vsini{}. A black dashed line indicates the colour adopted to split the main sequence into a red and blue part for magnitudes $18 < m_{\rm F438W} < 20$. On the right side, we zoom into the MSTO region. Highlighted in green are stars that are either likely members of NGC~1850B (circles), Be stars (squares), or shell stars (diamonds).}
    \label{fig:spexxy_results}
\end{figure*}

We studied the stellar rotation of the NGC~1850 stars in two ways. First, we looked at individual stellar spectra.  This analysis, which is described in Sect.~\ref{sec:rotation:individual}, itself rests on two pillars.  On the one hand, the spectral fits mentioned in Sect.~\ref{sec:analysis:spectroscopy} above, which provided us with a value for the (Gaussian) line broadening required to match the spectra. We converted this value into a \vsini{} measurement as described in Appendix~\ref{app:sigma2vsini}.  On the other hand, for stars that are hot enough to show \ion{He}{} lines in their spectra, we also obtained \vsini{} measurements by directly fitting \ion{He}{i} and \ion{He}{ii} lines covered by the MUSE spectral range.  We note that while there are four \ion{He}{i} lines available (at $4\,922~\text{\AA}$, $5\,016~\text{\AA}$, $6\,678~\text{\AA}$, and $7\,065~\text{\AA}$), we only used the reddest two lines, as the bluer lines are blended with \ion{Fe}{ii} lines (cf. Sec.~\ref{sec:rotation:combined}). While in Be stars of early spectral type, the \ion{He}{i} lines often show emission-line components, we do not expect such complications for the cooler stars residing in NGC~1850. The only \ion{He}{ii} line available is at $5\,411~\text{\AA}$. However, only four stars in our sample are hot enough to show \ion{He}{ii} absorption, and all of them belong to NGC~1850B according to the criterion of Sec.~\ref{sec:membership:ngc1850b}.

The second way in which we investigated stellar rotation was based on stacked spectra, obtained by summing up the MUSE spectra extracted for stars that are expected to be fast or slow rotators based on their positions in the HST colour-magnitude diagram.  We present this analysis in Sect.~\ref{sec:rotation:combined} below.

\subsection{Individual stars}
\label{sec:rotation:individual}

\begin{figure}
    \centering
    \includegraphics[width=.95\linewidth]{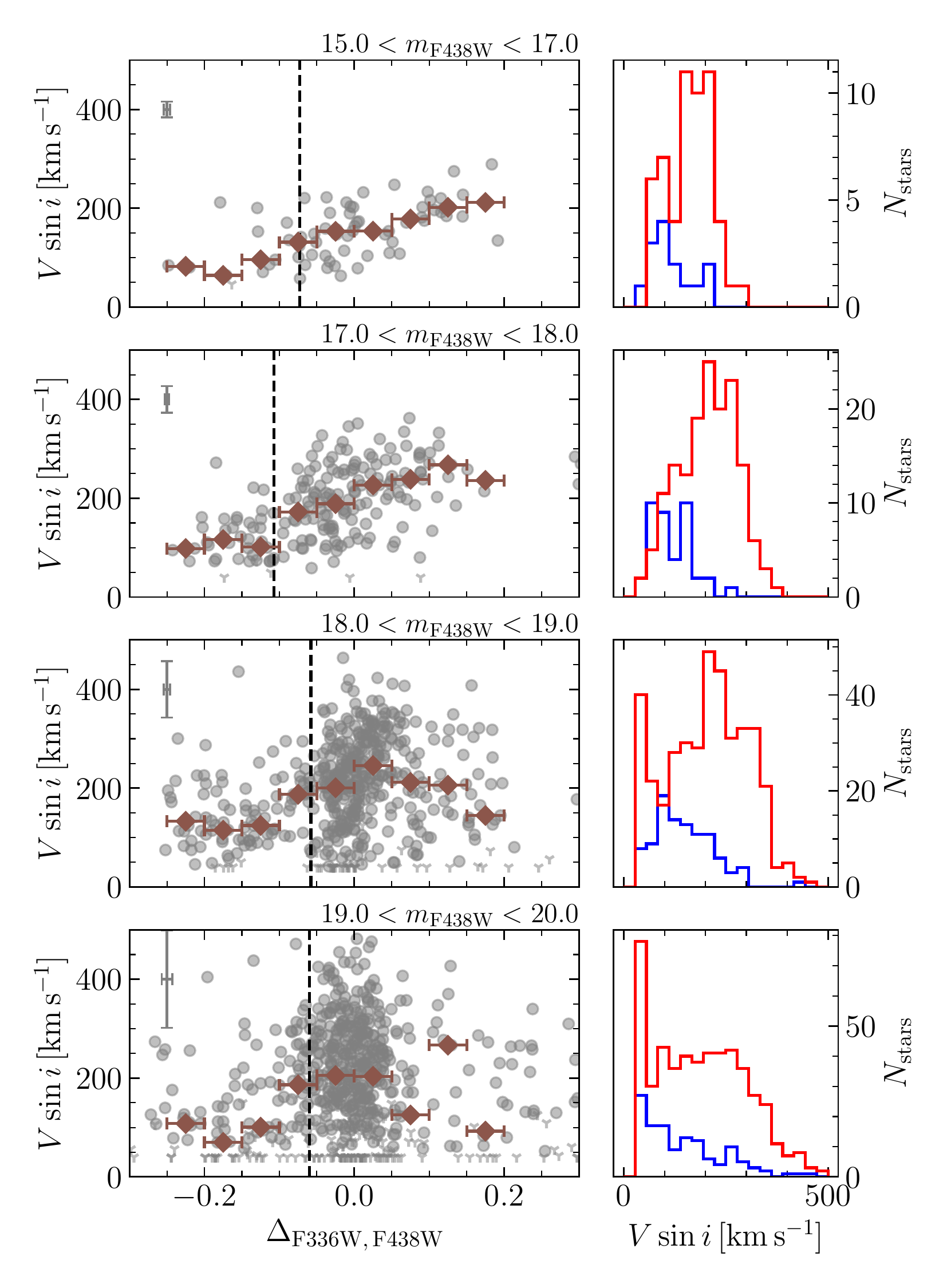}
    \caption{The left panels show, for four different $m_{\rm F438W}$ magnitude bins, the measured \vsini{} of main sequence stars as a function of pseudo-colour, i.e. the colour offset of a star relative to the median colour of the main sequence stars at a given magnitude. Grey points represent \vsini{} measurements for individual stars, with upper limits shown as triangles, while brown diamonds indicate median \vsini{} measurements (including upper limits) in pseudo-colour bins of 0.05~mag width. The average uncertainties per magnitude bin are shown in the top left of each panel. Vertical black dashed lines indicate the location of the 20th percentile in pseudo-colour, with 20\% roughly corresponding to the ratio of blue main sequence stars determined photometrically. In the right panels, we show histograms of the \vsini{} distributions for each bin, separated according to the 20th percentile divisions included in the left bins.}
    \label{fig:vsini_distributions}
\end{figure}

As our parent sample for obtaining individual \vsini{} measurements, we considered the same set of \nstars{} spectra mentioned in Sect.~\ref{sec:stellar_parameters} above. For each spectrum in the parent sample, the fitting of individual lines was performed as outlined below. However, the \textsc{Spexxy} sample was cleaned from any Be stars or likely members of NGC~1850B. Be stars, which show hydrogen line emission that are likely to impact the spectral fitting, will be discussed in Sec.~\ref{sec:be_and_shell}. NGC~1850B members were discarded because our isochrone fitting was not matched to the young age of this cluster. Excluding these two types of stars left us with a sample of 1\,873 stars with \vsini{} measurements based on the \textsc{Spexxy} fits. Note that both Be stars and NGC~1850B members were not removed from the line-fitting sample, as their \ion{He}{} lines may still result in useful \vsini{} measurements.

When performing the line fitting, each \ion{He}{} line was fitted individually.  Prior to the fits, the continuum of each spectrum was determined via a polynomial fit (from which the spectral lines were iteratively excluded using kappa-sigma clipping) and used to normalise the spectra.  The fits utilise the line spread function (LSF) of MUSE at the wavelength of each line.  This is possible because the MUSE LSF as a function of wavelength is measured by the pipeline and provided in the LSF\_PROFILE calibration file \citep[see Sect. 4.10 in][]{muse_pipeline}. We performed a non-linear least squares fit in which the initial LSF profile was broadened with a kernel accounting for stellar rotation with a given \vsini{}. To calculate the latter, we used the \textit{rotBroad} function available in the \textsc{PyAstronomy} package \footnote{https://github.com/sczesla/PyAstronomy} package \citep{pya}, which determines the impact of stellar rotation using the prescriptions provided in \citet{gray2008}.  The fits were carried out using \textsc{lmfit} \citep{lmfit} with a Levenberg-Marquardt optimisation and during each fit, we varied \vsini{}, the line centre $\lambda_{\rm c}$, and the total line flux $f$.  We only considered a line as successfully fitted if $\lambda_{\rm c}$ was within $0.5\,\text{\AA}$ of the tabulated value and the value of $f$ exceeded the noise level of the continuum around the line by a factor of $7$. If more than one line was successfully fitted per spectrum, the results from individual line fits were averaged. Furthermore, we used such cases to calibrate our measurement uncertainties. To this aim, the difference between each pair of \vsini{} values derived from the same spectrum was obtained and normalised by the squared sum of their uncertainties. In the case of correctly calibrated uncertainties, the resulting distribution should be Gaussian with a standard deviation of unity. Otherwise, the calibration is performed via multiplication of the uncertainties with the actual standard deviation of the distribution. We found that the uncertainties returned by \textsc{lmfit} underestimated the true uncertainties by a factor of $1.5$.

We obtained a sample of 306 stars with line-based \vsini{} values.  The considerably lower number of stars compared to the \textsc{Spexxy} approach can be explained by the gradual disappearance of the \ion{He}{} lines at lower effective temperatures.  As a consequence, the faintest stars for which we can still analyse the \ion{He}{} lines are at $m_{\rm F438W}\sim18.7$, corresponding to $T_{\rm eff}\sim13\,000~{\rm K}$ in the \textsc{Spexxy} fits and the isochrones.

Where \vsini{} measurements from both methods were available for the same stars, the two measurements were averaged, resulting in a final sample of 1\,963 stars with individual \vsini{} results available. The remaining 221 stars from the parent sample were either Be or NGC~1850B stars for which the \vsini{} measurement via \ion{He}{} lines failed. Where possible, we compared the results from the two methods and found that on average, the \vsini{} values derived using \textsc{Spexxy} where larger than those derived from the single-line fits by $16~{\rm km\,s^{-1}}$. The standard deviation between the results from the two methods is $50~{\rm km\,s^{-1}}$, which is in agreement with the expected accuracy of our \vsini{} measurements, taking into account the spectral resolution and wavelength coverage of MUSE as well as the added complication of the nebulosity impacting the \textsc{Spexxy} fits in the Balmer lines.

In Fig.~\ref{fig:spexxy_results}, we show the measured \vsini{} values for stars within NGC~1850 in a $m_{\rm F336W}-m_{\rm F438W}$ vs. $m_{\rm F438W}$ CMD, focusing on the main sequence (MS) and main sequence turn-off (MSTO). Overall, we observe that for a given magnitude there is a correlation between colour and \vsini{}, in the sense that redder stars rotate faster. To illustrate this, we show in Fig.~\ref{fig:vsini_distributions} the colour dependence of our \vsini{} measurements for different magnitude bins. For this purpose, we define a colour distance or ``pseudo-colour'' $\Delta_{\rm F336W,\,F438W}$ which measures the horizontal difference of a star relative to the median colour of main sequence stars at a given $m_{\rm F438W}$ magnitude. Besides the individual values, we also show in the left panels of Fig.~\ref{fig:vsini_distributions} the median \vsini{} values in pseudo-colour bins of 0.05~mag as brown diamonds. For each magnitude bin, we finally show histograms of the \vsini{} measurements in the right panels, separately for the bluest 20\% of the stars and the remaining 80\%. This division is motivated by the photometric analysis, which resulted in a fraction of 20\% of blue main sequence stars (cf. Sec.~\ref{sec:analysis:photometry}).

The \vsini{} distributions shown in the upper two rows of Fig.~\ref{fig:vsini_distributions} look quite similar to those obtained for the MSTO of the 1.5~Gyr old cluster NGC~419 in \citet{kamann_ngc1846}. They confirm that indeed, stellar rotation plays a dominant role in shaping the MSTO of YMCs.

It is interesting to note that while for the brightest bin ($m_{\rm F438W}<17$), the relation between \vsini{} and pseudo-colour appears continuous, a drop starts to appear for the other magnitude bins, in the sense that the bluest $\sim$20\% of the stars have substantially lower \vsini{} values on average than the remaining stars. In particular at fainter magnitudes ($18<m_{\rm F438W}<20$), where the two main sequences can be most easily distinguished, a clear bimodality is visible, with the first three bins (with $\Delta_{\rm F336W,\,F438W}<-0.1$) having median \vsini{} values of $100-120\,{\rm km\,s^{-1}}$, while the following bins have median \vsini{} values of $190-220\,{\rm km\,s^{-1}}$. This confirms previous suggestions that the split main sequence in young ($<300$~Myr) clusters is primarily due to stellar rotation \citep[e.g.,][]{marino18_m11}. Based on the lower three panels of Fig.~\ref{fig:vsini_distributions}, we adopt median \vsini{} values of $110\pm20\,{\rm km\,s^{-1}}$ and $210\pm20\,{\rm km\,s^{-1}}$ for the blue and red main sequences, respectively.  The uncertainties that we assign to the median values are based on the scatter between the values of the individual pseudo-colour bins and the mean difference between the line-based and \textsc{Spexxy}-based \vsini{} values reported above, which we consider as representative for the strengths of the systematic errors involved in our analysis.

As will be discussed further in Sec.~\ref{sec:rotation:models}, where we compare our measurements to stellar evolutionary models, the critical (break-up) velocity $V_{\rm crit}$ for the stars included in Fig.~\ref{fig:vsini_distributions} is $\sim450\,{\rm km\,s^{-1}}$ ($\pm50\,{\rm km\,s^{-1}}$, depending on mass). We note that a small fraction of the measurements (23 stars) shown in Fig.\ref{fig:vsini_distributions} exceed this value. While some of these outliers can be explained by the limited accuracy of our measurements for the faintest stars in the sample (see error bars included in the left panels of Fig.~\ref{fig:vsini_distributions}), visual inspection of the spectra of some of these sources also reveals SB2 binaries (i.e. binaries with two luminous companions that both contribute lines to the combined spectrum). These spectra have multiple component absorption lines, meaning that the integrated profiles will be broader than expected for a single star, which results in the assignment of high \vsini{} values for these sources. Given the small number of stars with \vsini$>V_{\rm crit}$, we did not make an effort to remove them from the computation of the median values included in Fig.~\ref{fig:vsini_distributions}.

There is still a possibility, however, that the \vsini{} distribution shown in Fig.~\ref{fig:vsini_distributions} is skewed by SB2 binaries.  However, only binary stars with a mass ratio $\sim 1$ and orbital periods $\lesssim 100~{\rm d}$ will produce combined spectra with a period-induced line broadening that is comparable to the observed \vsini{} values. Such binaries are expected to be rare, so that we do not expect a significant impact of SB2 binaries on our measured \vsini{} distribution. We note that for stars on the binary main sequence identified in Sec.~\ref{sec:analysis:photometry}, which roughly corresponds to the range in $\Delta_{\rm F336W,\,F438W}\gtrsim0.05$ in the lower two panels of Fig.~\ref{fig:vsini_distributions}, we do not observe a trend towards higher \vsini{} values that would hint towards an impact of binary orbital motions on the observed line profiles. On the contrary, we observe a decline in the median \vsini{} values (albeit with substantial scatter). This may be expected if the binary main sequence is mainly populated by binaries on relatively wide orbits ($\gtrsim 100~{\rm d}$) that have been (partially) braked by tidal interactions, as tidal interactions increase with increasing companion masses (and hence towards larger $\Delta_{\rm F336W,\,F438W}$). Indeed, the stars in binaries with orbits $\lesssim500\,{\rm d}$ are expected to be slowly rotating due to tidal interactions \citep{abt04}. In that case, however, the individual binary components would be blue main sequence stars. Their combination in a binary would result in a redder colour and therefore push the binary towards the red main sequence, rather than onto the binary main sequence. Hence, it is not surprising that most of the data points with \vsini{}$>V_{\rm crit}$ in Fig.~\ref{fig:vsini_distributions} have pseudo-colours similar to the red main-sequence stars. We note that in our analysis, we treated systems on the binary main sequence in the same way as all other identified cluster members. Given the expected small impact of their orbital motions on the observed line profiles, this approach appears justified in hindsight.

In Fig.~\ref{fig:spexxy_results}, it is further visible that the region of the CMD populated by blue stragglers (with $m_{\rm F336W} - m_{\rm F438W} \lesssim -1$ and $m_{\rm F438W} \lesssim 18.5$) hosts stars with a large range in \vsini{} values. However, care must be taken as this region is also populated by main sequence stars of NGC~1850B. We highlight members of the younger cluster (according to Sect.~\ref{sec:membership:ngc1850b}) via green circles in Fig.~\ref{fig:spexxy_results}.  When we omit the likely NGC~1850B members, we find that a majority of the blue stragglers are relatively slow rotators (with \vsini{}$\lesssim150\,{\rm km\,s^{-1}}$), but that some blue stragglers with $\vsini\gtrsim200\,{\rm km\,s^{-1}}$ are also observed. The latter include a spectroscopically identified Be star, visible in the right panel of Fig.~\ref{fig:spexxy_results} at $m_{\rm F336W} - m_{\rm F438W}=-1.07$ and $m_{\rm F438W}=17.1$ (as well as several Be star candidates identified photometrically, cf. Sec.~\ref{sec:be_stars}). \citet{wang2020,wang2022} predict that the blue straggler region should be predominantly populated by slow-rotating merger products. In this scenario, the fast rotators might be considered as products of recent mergers that did not yet have time to spin down -- and their number could be used to estimate spin down times. However, besides mergers, mass transfer provides an alternative pathway to blue straggler formation, in which case a fast rotating product is expected. Along these lines, it is interesting to note that the three blue stragglers with \vsini{}$>300\,{\rm km\,s^{-1}}$ all show evidence for radial velocity variations in the MUSE data. The binary properties of our sample will be the topic of a separate publication (Saracino et~al., in prep.). Given that many Be stars have been found to reside in binary systems with stripped companions, the presence of Be stars among the blue stragglers of NGC~1850 seems unsurprising.

We will further discuss the implications of the results shown in Figs~\ref{fig:spexxy_results} and \ref{fig:vsini_distributions} below in Sect.~\ref{sec:rotation:models}, following a summary of the results derived from the combined spectra. The rotation rates of Be and shell stars, which are highlighted in the right panel of Fig.~\ref{fig:spexxy_results}, will be discussed in Sec.~\ref{sec:be_and_shell}.

\subsection{Analysis of combined spectra}
\label{sec:rotation:combined}

\begin{figure}
    \centering
    \includegraphics[width=.95\linewidth]{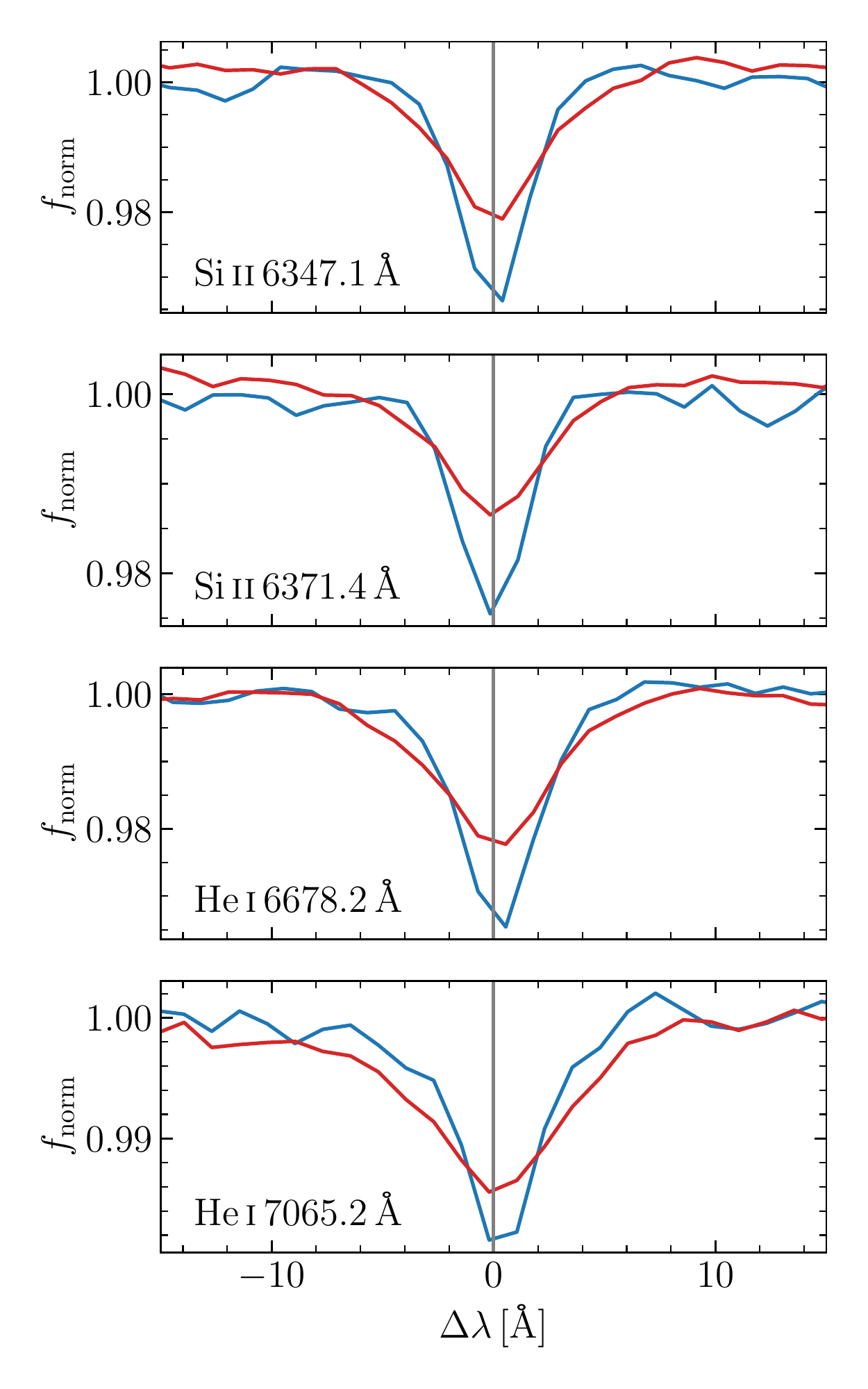}
    \caption{Isolated helium and silicon lines identified in the combined spectra created for the red and blue main sequence stars. In each panel, we provide the normalized line profiles for both spectra. The line identification and central wavelength of each line are provided in the lower left corner.}
    \label{fig:vsini_lines}
\end{figure}

In total, our MUSE spectroscopic sample contains 337 blue main sequence stars and 988 red main sequence stars with S/N>10.\footnote{This corresponds to a fraction of $\sim$27\% blue main sequence stars, higher than the 20\% we assumed in Sec.~\ref{sec:analysis:photometry}. The reason for this difference are the different spatial distributions of the two populations in combination with the complicated spectroscopic selection function (as S/N depends on both brightness and location of a star).} We made further use of the large sample sizes by creating S/N-weighted mean spectra for both populations, with the aim of detecting the same \ion{He}{i} lines we were able to identify in the individual spectra of turn-offs stars, or any metallic lines that are sufficiently narrow to enable direct measurements of \vsini{} via single-line fits. Indeed, the combined spectra showed a number of such lines, namely the \ion{He}{i} lines at $4\,922\,\text{\AA}$, $5\,016\,\text{\AA}$, $6\,678\,\text{\AA}$, and $7\,065\,\text{\AA}$, as well as \ion{Si}{ii} lines at $6\,347\,\text{\AA}$, $6\,371\,\text{\AA}$, and the \ion{O}{i} line at $7\,774\,\text{\AA}$. In order to verify whether these lines are truly isolated at the spectral resolution of MUSE in the $T_{\rm eff}$ range under consideration, we downloaded a spectrum of the star HD~196426 from the 2nd data release of the X-Shooter spectral library \citep[XSL,][]{gonneau2020}. Based on the spectral analysis performed by \citet{arentsen2019}, the star has similar stellar parameters compared to the average main sequence star entering our combined spectra. We found that the two \ion{He}{i} lines at $4\,922\,\text{\AA}$, $5\,016\,\text{\AA}$ are blended with \ion{Fe}{ii} lines (at $4\,924\,\text{\AA}$ and $5\,018,\text{\AA}$, respectively), while the \ion{O}{i} line at $7\,774\,\text{\AA}$ is a doublet. Hence we are left with four truly isolated lines, for which the line profiles for the two combined spectra are compared in Fig.~\ref{fig:vsini_lines}. In all cases, it is clearly visible that the line profile in the spectrum of the red main sequence stars is broader compared to the spectrum of the blue main sequence stars, confirming the results from the individual stars of a strong difference in \vsini{} between the two sequences. We note that the difference in effective temperature between the red and blue main sequence stars ($\sim1\,000~{\rm K}$) are very unlikely to be responsible for the observed differences, given that its impact on the lines shapes is below the resolving power of MUSE and that, as we will show below, we obtain consistent results for different elements (\ion{He}{} and \ion{Si}{}). In addition, note that the red main sequence stars are \emph{cooler} than the blue main sequence stars. Therefore, temperature-dependent effects would counteract the observed differences.

In order to quantify the line width differences visible in Fig.~\ref{fig:vsini_lines}, we carried out single-line fits in a very similar fashion to those described for the individual spectra in Sec.~\ref{sec:rotation:individual}. Again, we operated on normalized spectra and fitted for \vsini{} by convolving a model for the MUSE LSF with a broadening kernel generated using the \textit{rotBroad} function available in the \textsc{PyAstronomy} package \citep{pya}. The results are summarized in Table~\ref{tab:line_fit_results}. The uncertainties included in Table~\ref{tab:line_fit_results} were obtained by repeating the fits for 100 bootstrap realizations, in which we randomly picked red and blue main sequence stars from our sample and averaged their spectra in the same manner as before.

The weighted averages of the results listed in Table~\ref{tab:line_fit_results} are $99\pm5\,{\rm km\,s^{-1}}$ and $188\pm7\,{\rm km\,s^{-1}}$, respectively, for the blue and red main sequences. The results are in very good agreement with those obtained from the analyses of the individual spectra, giving us further confidence into the validity of our approach.

We performed another check in which we took the XSL spectrum of HD~196426, for which \citet{arentsen2019} provide a low \vsini{} of $20\,{\rm km\,s^{-1}}$, and converted it into a mock MUSE spectrum. This was achieved by convolving the XSL spectrum with the MUSE LSF model provided by the pipeline as a function of wavelength, adding Gaussian noise such that we achieved similar S/N as in the combined MUSE spectra, and finally rebinning the spectrum to a constant sampling of $1.25\,\text{\AA}$ per pixel. For the resulting mock spectrum, we performed the same line fits as for the two combined spectra. For all four lines listed in Table~\ref{tab:line_fit_results}, we found low \vsini{} values $\lesssim 40\,{\rm km\,s^{-1}}$, consistent with the detection threshold we expect in our data. This test shows that stellar rotation is indeed the dominant line broadening mechanism in the high \vsini{} regime we are concerned with.

\begin{table}
    \centering
    \caption{$V\sin i$ measurements obtained from single-line fits in the combined spectra.}
    \label{tab:line_fit_results}
\begin{tabular}{lcrrrr}
\hline
{} & {} & \multicolumn{2}{l}{red} & \multicolumn{2}{l}{blue} \\
$\lambda$ & {ion} & \vsini{} & $\epsilon_{\rm v\sin i}$ & \vsini{} & $\epsilon_{\rm v\sin i}$ \\ \hline
$\text{\AA}$ &  & $\rm km\,s^{-1}$ & $\rm km\,s^{-1}$ & $\rm km\,s^{-1}$ & $\rm km\,s^{-1}$ \\
\hline
6347.1 & \ion{Si}{ii} & 181 &         9 &    98 &         6 \\
6371.4 & \ion{Si}{ii} & 172 &        16 &    89 &        12 \\
6678.2 & \ion{He}{i}  & 188 &        12 &   118 &        21 \\
7065.2 & \ion{He}{i}  & 203 &        27 &   107 &        15 \\
\hline
\textbf{mean} &  & \textbf{188} & \textbf{7} & \textbf{99} & \textbf{5} \\
\hline
\end{tabular}
\end{table}

\subsection{Comparison to models}
\label{sec:rotation:models}

In Sec.~\ref{sec:rotation:individual} and \ref{sec:rotation:combined}, we found strong evidence for very different \vsini{} values along the blue and red main sequences of NGC~1850. In order to compare our results to the latest stellar evolutionary models, we need to correct the former for the effects of inclination. In this work, we assume a an isotropic distribution of spin axes, i.e. $\langle\sin i\rangle=\pi/4$. While it has been proposed that in star clusters the spin axes could be aligned as a result of cluster formation \citep{reyraposo18}, observational evidence for anisotropic spin distributions in clusters is still sparse \citep[e.g.,][]{lim19,healy21}. We did not make an effort to search for a possible deviation from isotropy in NGC~1850, but consider this a worthwhile endeveour for a future publication.

Under the assumption that the spin axes are distributed isotropically, our median values derived from the combined spectra correspond to mean equatorial velocities of $\langle V_{\rm surf,\,blue}\rangle=127\pm7\,{\rm km\,s^{-1}}$ and $\langle V_{\rm surf,\,red}\rangle=233\pm9\,{\rm km\,s^{-1}}$.  Using the values from the individual fits instead, we find slightly higher values of $\langle V_{\rm surf,\,blue}\rangle=140\pm26\,{\rm km\,s^{-1}}$ and $\langle V_{\rm surf,\,red}\rangle=276\pm26\,{\rm km\,s^{-1}}$.  

\begin{figure}
    \centering
    \includegraphics[width=.95\linewidth]{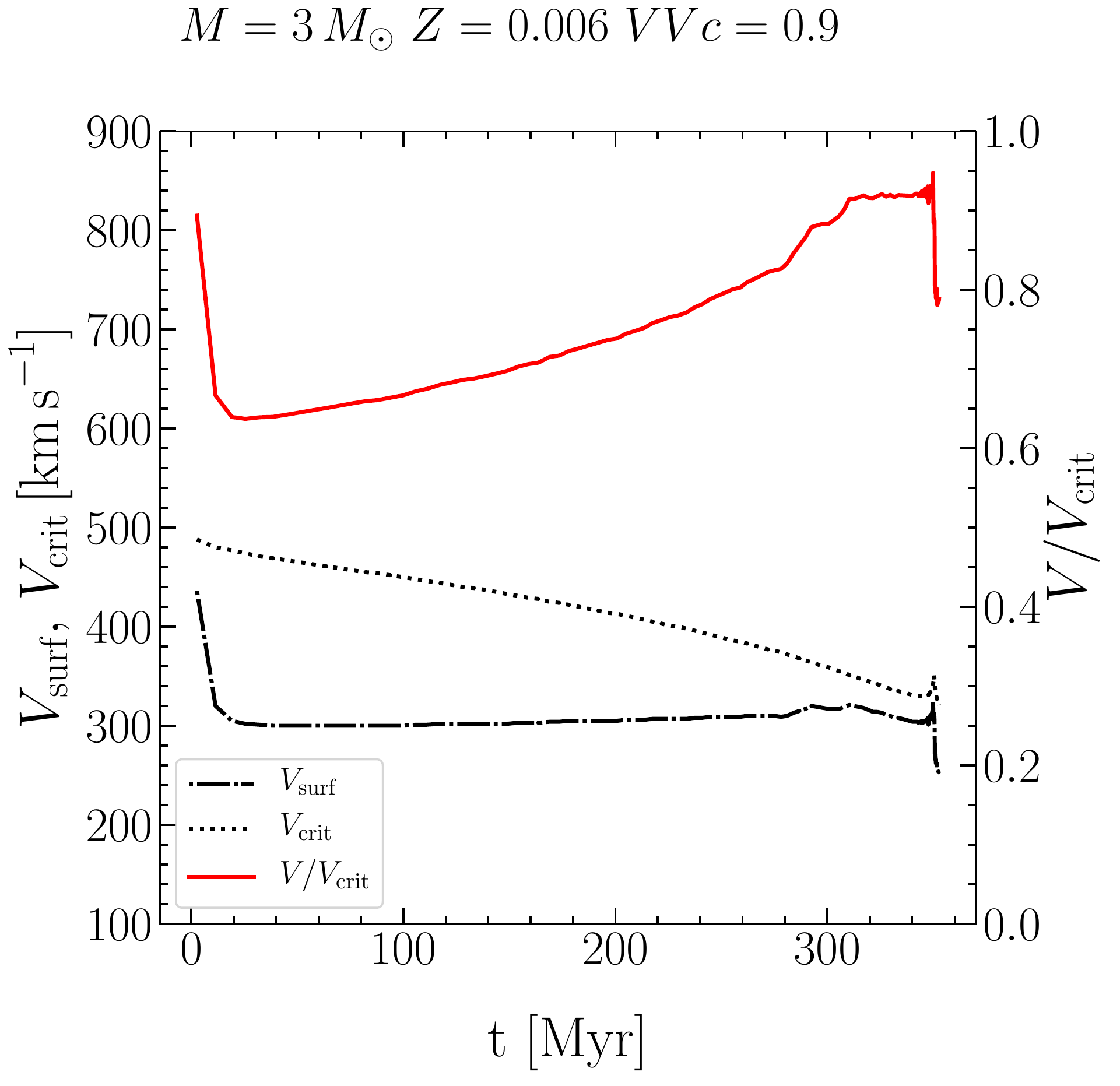}
    \caption{The time evolution of the surface velocity $V_{\rm surf}$ (dash-dotted line), the critical velocity $V_{\rm crit}$ (dotted line), and the ratio $V/V_{\rm crit}$ (solid red line) as a function of age in the SYCLIST models. Shown are the predictions for a star of $3\,M_{\odot}$ with a metallicity of $Z=0.006$ and an initial $V/V_{\rm crit}=0.9$.}
    \label{fig:syclist}
\end{figure}

For the magnitude range $18.0 < m_{\rm F38W} < 20.0$ and the adopted cluster properties of NGC~1850, the MIST isochrones predict stellar masses between $2.3\,M_{\rm \odot}$ and $4.2\,{\rm M_\odot}$. The median predicted masses for our samples of red and blue main sequence stars are $3.01\,{\rm M_\odot}$ and $3.07\,{\rm M_\odot}$, respectively. As mentioned above, the SYCLIST models published in \citet{georgy2013} predict a critical velocity of $V_{\rm crit}=450\,{\rm km\,s^{-1}}$ for a $3\,{M_\odot}$ star at the age and metallicity of NGC~1850 (cf. Fig.~\ref{fig:syclist}). Adopting this value, our median velocities correspond to ranges in $V/V_{\rm crit}$ of $25-30\%$ and $50-60\%$ for the blue and red main sequences, respectively. To enable a better comparison with isochrone predictions, we also express our results in terms of the critical angular velocity, $\Omega_{\rm crit}$. Note that the relationship between $V/V_{\rm crit}$ and $\Omega/\Omega_{\rm crit}$ is non-linear, given the deformation of a star as its spin increases \citep[e.g.,][]{granada2013}. We find ratios of $35-40\%$ and $67-79\%$ for the median $\Omega/\Omega_{\rm crit}$ of the two main sequences.

Our results suggest that rotation velocities close to the critical value\footnote{Our definition of ``near critical'' or ``close to critical'' is $\gtrsim80\%$ of the break-up velocity} are not required in order to explain the split main sequence. At face value, this is in agreement with the latest predictions from isochrone fitting: The MIST models discussed in Sect.~\ref{sec:analysis:isochrones} suggest that blue main sequence stars have $\Omega/\Omega_{\rm crit}\lesssim0.3$, whereas red main sequence stars have $\Omega/\Omega_{\rm crit}\gtrsim0.6$. The models presented by \citet{wang2020} are instead parametrized in $V/V_{\rm crit}$ and \citet{wang2022} found that values of 0.35 and 0.65 match the observed main sequences in the young massive cluster NGC~1755.  However, one must be careful in that both of the aforementioned models use the critical velocity at the zero-age main sequence (ZAMS). As shown by \citet{hastings20}, $V/V_{\rm crit}$ (and equivalently $\Omega/\Omega_{\rm crit}$) can vary substantially over the main-sequence lifetime of a star. To illustrate this, we show in Fig.~\ref{fig:syclist} the time evolution of the surface velocity, its critical value, and the ratio of two as predicted by the SYCLIST models by \citet{georgy2013} for a $3\,{\rm M_\odot}$ star in NGC~1850 \citep[see][for similar plots showing the predictions for the models used by \citealt{wang2022}]{hastings20}. Even though the model was initialized with a $V/V_{\rm crit}=0.9$ at the ZAMS, its value at the age of NGC~1850 (100~{\rm Myr}) has decreased to $\sim0.7$, as a result of a steep drop in surface velocity in the first $10~{\rm Myr}$.

In Fig.~\ref{fig:syclist}, it can be seen that apart from the initial drop, the SYCLIST models predict the surface velocity to barely change during the main sequence evolution \citep[see also][]{bastian20}.  On the other hand, a substantial decrease in the critical velocity, from an initial value $\sim500\,{\rm km\,s^{-1}}$ at the zero-age main sequence to $\sim320\,{\rm km\,s^{-1}}$ at the turn-off age of roughly $300\,{\rm Myr}$, is predicted. This is caused by the expansion of the stellar envelope in response to a strong chemical gradient between the convective core and the radiative envelope. Comparing the value at the MSTO to the \vsini{} distribution of red main sequence stars in Fig.~\ref{fig:vsini_distributions} suggests that a substantial fraction of the stars will be close to critically rotating when reaching the end of their main sequence lifetimes.

Interestingly, our data do not show an increase in $V/V_{\rm crit}$ when approaching the MSTO. For an age of $100\,{\rm Myr}$ and $Z=0.006$, the $V_{\rm crit}$ predicted by the SYCLIST models for stars of $4\,{\rm M_\odot}$ and $5\,{\rm M_\odot}$ are $\sim450\,{\rm km\,s^{-1}}$ and $\sim400\,{\rm km\,s^{-1}}$, respectively. As $5\,{\rm M_\odot}$ corresponds to the turn-off mass of NGC~1850, we can compare the latter value to the \vsini{} measurements shown in the top row of Fig.~\ref{fig:vsini_distributions}. Barely any values exceed $250\,{\rm km\,s^{-1}}$, so that, unless the observed distribution is significantly compressed by inclination effects, the bulk of our MSTO sample is restricted to $V/V_{\rm crit}\lesssim0.6$. At face value, this could be taken as evidence for a lack of near-critically rotating stars at the MSTO of NGC~1850.  However, the \vsini{} measurement for (almost) critically rotating stars can be biased towards lower values because of the rotationally induced darkening of the equatorial regions of the star \citep{townsend04}. In addition, NGC~1850 contains a large fraction of Be stars among its MSTO population (cf. Sec.~\ref{sec:be_and_shell}), which are considered to be the outcome of near-critical stellar rotation.

Comparing the upper two rows in Fig.~\ref{fig:vsini_distributions}, we observe a shift towards lower \vsini{} values for the brightest stars. We found a similar trend in our previous work on the 1.5~Gyr old cluster NGC~1846 \citep{kamann_ngc1846}. This might indicate that the stars at the tip of the main sequence are already being braked as they start evolving into red giants.

Fig.~\ref{fig:spexxy_results} shows that the trend that at a given magnitude, bluer stars rotate slower than redder stars is not restricted to the magnitude range of the split main sequence, but persists all the way to the MSTO. When comparing to the isochrone tracks shown in Fig.~\ref{fig:cmd}, it becomes evident that this \vsini{} dependency of the colour is only predicted for the fainter stars in our sample (i.e. in the magnitude range showing a split main sequence), yet not at the MSTO, where the isochrone tracks for different rotation rates merge. Interestingly, this is not the case in the models presented by \citet{wang2022}, where a \vsini{} dependency of the observed colour is predicted even for MSTO stars. This illustrates how our observations can be used to scrutinise predictions from stellar evolutionary models.

Overall, our observations appear to be in reasonable agreement with current stellar evolutionary model predictions. In particular, the predicted difference in the rotation rates of blue and red main sequence stars is confirmed. However, an in-depth comparison is hampered by the different reference times. While observations naturally reveal the stellar rotation properties at a given cluster age, models typically refer to these properties at the zero-age main sequence. This is particularly problematic as different models predict different relations between rotation at the zero-age main sequence and at a given cluster age. For example, the drop in $V$ visible at early ages in Fig.~\ref{fig:syclist} appears to be absent in the models used by \citet{hastings20}. Model predictions of the surface velocity $V$ as a function of cluster age will offer a promising venue for future research, enabling the combined analyses of photometry and \vsini{} measurements in order to better understand star cluster populations \citep[e.g.,][]{lipatov22}.

\section{B\lowercase{e} and Shell Stars}
\label{sec:be_and_shell}

\begin{figure}
    \centering
    \includegraphics[width=.95\linewidth]{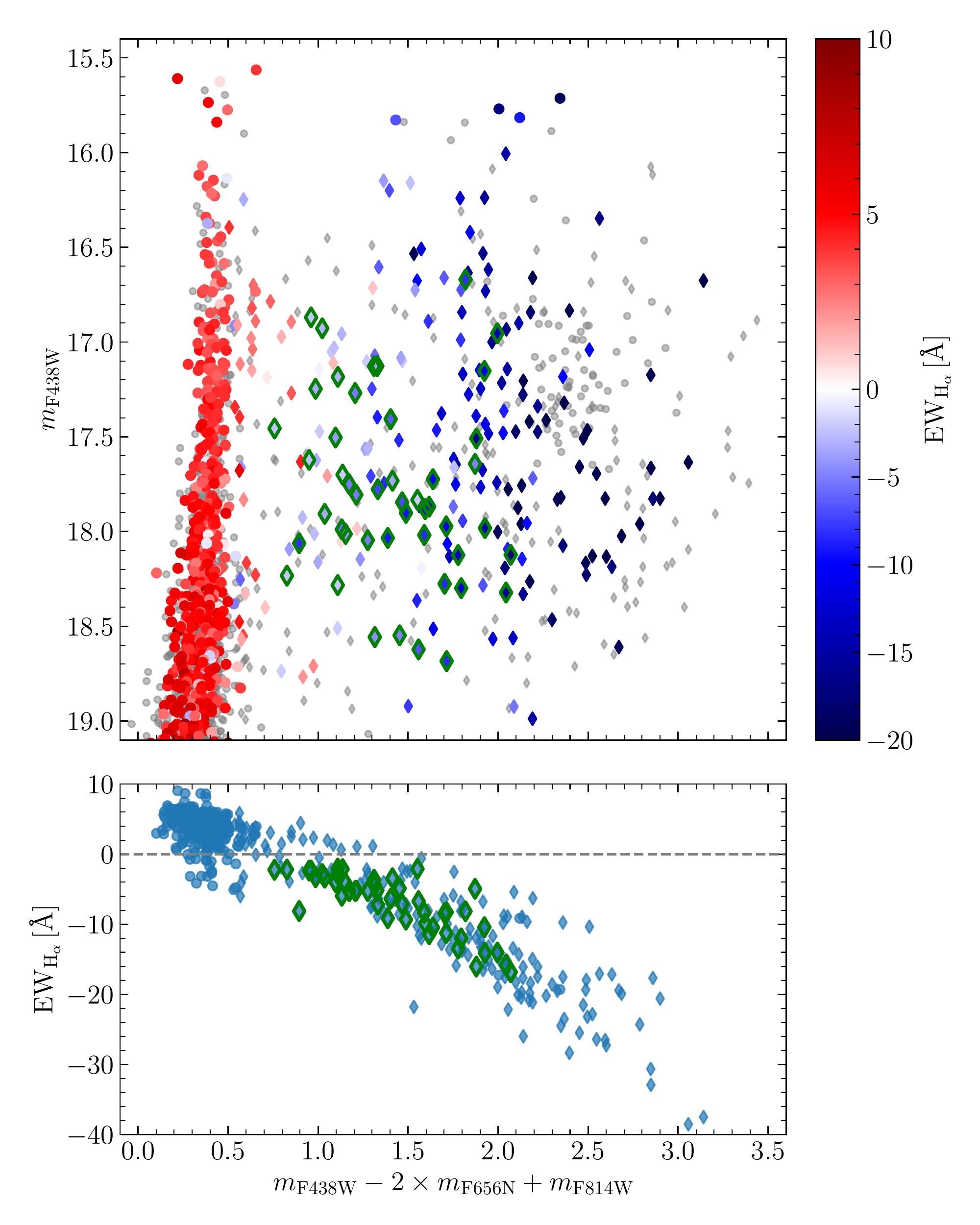}
    \caption{H$\alpha$ equivalent width as a function of the photometric index $m_{\rm f438W} - 2\times m_{\rm F656N} + m_{\rm F814W}$.  The upper panel displays the photometric index as a function of $m_{\rm F438W}$ colour. Stars for which MUSE spectra are available are colour-coded according to their H$\alpha$ equivalent width.  In the lower panel, we show the relation between H$\alpha$ equivalent width and $m_{\rm F438W} - 2\times m_{\rm F656N} + m_{\rm F814W}$.  In both panels, objects identified as shell stars (cf. Sec.~\ref{sec:shell_stars}) are highlighted using green frames. In both panels, stars photometrically identified as Be star candidates are shown as diamonds, while all other stars are shown as circles.}
    \label{fig:halpha_cmd}
\end{figure}

As established above, NGC~1850 hosts a substantial population of stars that are rapidly rotating and previous studies have found large populations of Be stars within the cluster using HST $m_{\rm F656N}$ narrowband photometry \citep[][]{bastian17,milone18_be}.  A common method to identify Be stars photometrically is by detecting outliers in a colour calculated from $m_{\rm F656N}$, which is centred on H$\alpha$, and a nearby broadband filter.  As our photometry is lacking any $V$-band equivalent and using a bluer (redder) broadband filter could result in cool (hot) stars being misclassified as Be, we calculated the photometric index $C_{\rm 438,\,656,\,814}=m_{\rm F438W}-2\times m_{\rm F656N}+m_{\rm F814W}$ and plotted it as a function of $m_{\rm F438W}$ (cf. Fig.~\ref{fig:halpha_cmd}).  For the magnitude range $16<m_{\rm F438W}<19$, we defined the ridgeline of the main sequence (after removing all stars with $C_{\rm 438,\,656,\,814}>0.5$) and selected as Be stars all sources that deviated from the ridge line by more than $6\times$ their uncertainty in $C_{\rm 438,\,656,\,814}$.\footnote{lower thresholds would result in the misidentification of normal MSTO stars as Be stars, given the spread of the MSTO in $m_{\rm F438W}-m_{\rm F814W}$ colour.} This resulted in a sample of 397 Be star candidates. MUSE spectra are available for 218 of them, with most of the remaining candidates being located outside of the observed MUSE field of view.

In order to identify Be stars in the MUSE spectra, we extended the fitting of individual lines described in Sect.~\ref{sec:rotation:individual} to the H$\alpha$ line, using a double Gaussian profile in order to account for a potential emission component. Following the fits, we summed up the equivalent widths of the two Gaussian components. In Fig.~\ref{fig:halpha_cmd}, we show the resulting H$\alpha$ equivalent (${\rm EW}_{\rm H\alpha}$) width as a function of the photometric index $C_{\rm 438,\,656,\,814}$, with the lower panel showing an almost linear relation between the spectroscopic and photometric indices.

Fig.~\ref{fig:halpha_cmd} demonstrates that there is good agreement between the stars showing H$\alpha$ emission in the MUSE data and those showing $m_{\rm F656N}$ excess in the HST data.  We measured ${\rm EW}_{\rm H\alpha}<0$ in the spectra of 202 stars, out of which 185 have also been flagged as Be star candidates using the photometric approach. For another 33 photometric Be star candidates, we measured ${\rm EW}_{\rm H\alpha}>0$. As can be verified from Fig.~\ref{fig:halpha_cmd}, the latter stars have low $m_{\rm F656N}$ excesses, indicating that their emission components are not strong enough to fill in the entire H$\alpha$ lines. This was confirmed by visual inspection of some of the spectra.

The relation between ${\rm EW}_{\rm H\alpha}$ and $C_{\rm 438,\,656,\,814}$ shown in the bottom panel of Fig.~\ref{fig:halpha_cmd} does show significant scatter. Part of it can be explained by the time gap between the HST and MUSE observations of about 5~years, as Be stars show variability on smaller timescales \citep{labadie18}. In addition, some of the H${\rm \alpha}$ line profiles of the spectra underlying Fig.~\ref{fig:halpha_cmd} show contamination from the nebulosity in the field of view of the cluster (see Fig.~\ref{fig:muse}). We note that the nebulosity varies on scales similar to the resolution of MUSE ($\sim0.5$") and its H$\alpha$ emission line extends across seven wavelength bins (corresponding to 9~$\text{\AA}$ or $400\,{\rm km\,s^{-1}}$). Hence, even following a PSF-based extraction of the stellar spectra, a fraction of the stellar H$\alpha$ lines remained affected by the nebulosity.

\subsection{Be star demographics}
\label{sec:be_stars}

\begin{figure}
    \centering
    \includegraphics[width=.95\linewidth]{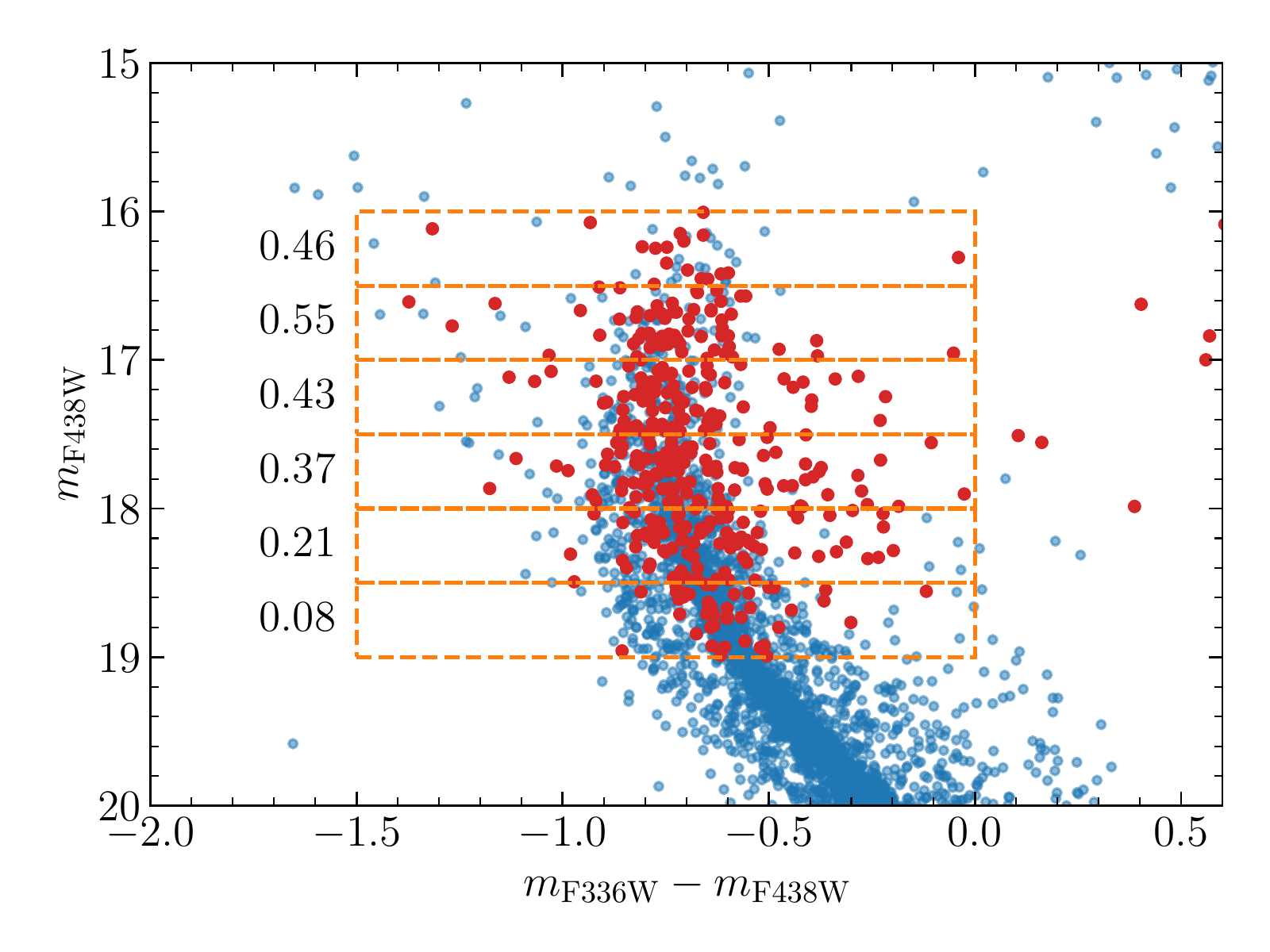}
    \caption{HST colour-magnitude diagram of likely cluster members of NGC~1850, with Be star candidates shown in red and all other clusters stars shown in blue. Dashed orange lines show the selection boxes of width $\Delta m_{\rm F814W}=0.5$ that were used to measure the fraction of Be stars relative to the total number of stars as a function of magnitude. The resulting Be fraction for each box is provided to its left.}
    \label{fig:be_frac}
\end{figure}

Fig.~\ref{fig:halpha_cmd} suggests a trend of a decreasing fraction of Be stars with increasing $m_{\rm F438W}$ magnitude.  This trend, which has been noted previously for NGC~1850 \citep[][]{bastian17} is unlikely to be caused by incompleteness due to decreasing emission line strength, even though the latter is generally expected to decrease for later spectral types given the lower ionizing fluxes. However, in the magnitude range $m_{\rm F438W}\gtrsim18.3$, the few Be stars that are observed still cover a similar range in $m_{\rm F656N}$ excess and $EW_{\rm H\alpha}$ as the more numerous brighter Be stars. Also, we do not expect the diffuse nebulosity to play a significant role here because otherwise, we would expect to see a significant increase in the width of the main sequence in the top panel of Fig.~\ref{fig:halpha_cmd} towards fainter magnitudes.

In order to have a closer look at the fraction of Be stars as a function of magnitude, we revert to the photometric selection. This is because out of the 397 Be star candidates identified based on their $m_{\rm F656N}$ excess, spectroscopy is only available for 218 stars.  Given the good agreement between the spectroscopic and photometric selection illustrated in Fig.~\ref{fig:halpha_cmd}, we do not expect any significant contamination of the Be star sample when using the latter.

In Fig.~\ref{fig:be_frac}, we show the position of the Be stars in a ($m_{\rm F336W} - m_{\rm F814W}$, $m_{\rm F814W}$) CMD.  In the magnitude range $16<m_{\rm F438W}<19$, we constructed six selection boxes of 0.5~mag width and determined the ratio of Be stars to all other cluster members for each box.  It can be seen from Fig.~\ref{fig:be_frac} that the fraction increases by an order of magnitude between $m_{\rm F438W}=19$ and $m_{\rm F438W}=17$. According to the MIST isochrones, the stellar mass increases from $3.2~{\rm M_\odot}$ to $4.8~{\rm M_\odot}$ in this range. We note that similar behaviour has been observed in a sample of LMC/SMC stellar clusters by \citet{milone18_be}, where few or no Be stars were found on the main sequence but large fractions were found near the MSTO.

Such a trend with magnitude is in agreement with the model predictions of \citet{hastings20}, who treat the Be phenomenon as being due primarily to single star evolution. In their models, the trend is explained by the dropping of the critical velocity of a star as it approaches the MSTO (cf. Fig.~\ref{fig:syclist}). Indeed, based on the results we obtained for the split main sequence, it seems plausible that a large fraction of the stars will be rotating close to their break-up velocities towards the end of their main sequence lifetimes. On the other hand, as discussed in Sec.~\ref{sec:rotation:models}, we do not find a significant number of near critical rotators among the current MSTO stars of NGC~1850. As shown in Sec.~\ref{sec:be_rotation} below, this finding also applies to the Be stars. While on average, they rotate faster than normal MSTO stars, their \vsini{} distribution does not extend to significantly higher values. In order to better understand the efficiency of forming Be stars via the single-star channel, it would be insightful to compare the \vsini{} distributions of main sequence stars with comparable masses of $3\,{\rm M_\odot}$ across clusters of different ages (up to $\sim$300~Myr) and look at the evolution of $V/V_{\rm crit}$.

A trend of Be star fraction with magnitude as shown in Fig.~\ref{fig:be_frac} can also be reproduced by models where Be stars are formed exclusively through binary channels \citep[][]{hastings21}.  However, such models require rather extreme assumptions regarding the initial binary fraction (near unity), a non-canonical initial mass function and very non-conservative mass transfer.

Note that the Be star fractions shown in Fig.~\ref{fig:be_frac} are based on the assumption that the Be star have comparable $m_{\rm F438W}$ magnitudes to normal B stars of the same mass. As shown by \citet{haubois12}, the presence of a disk can alter the broadband magnitudes of such stars, to an amount that depends on wavelength, disk viscosity, and inclination angle. \citet{haubois12} found differences between +0.2 and -0.35 magnitudes for the $V$ band. As the impact of the disk increases with wavelength, we expect no strong $m_{\rm F438W}$ magnitude differences between B and Be stars in NGC~1850.

We further looked at the spatial clustering of the Be stars and checked if they were more or less centrally concentrated than normal B stars of the same magnitude range in. When performing a two-sample Kolmogorov-Smirnov test on the distributions of projected distances to the cluster centre, we obtained a probability of 17\% that the two were drawn from the same parent distribution. One might argue that if a large fraction of Be stars live in binaries, they should be centrally concentrated compared to normal B stars. However, NGC~1850 is only $\sim$150~Myr old, so that barely any evolutionary mass segregation is expected.

\subsection{Shell stars}
\label{sec:shell_stars}

\begin{figure}
    \centering
    \includegraphics[width=.95\linewidth]{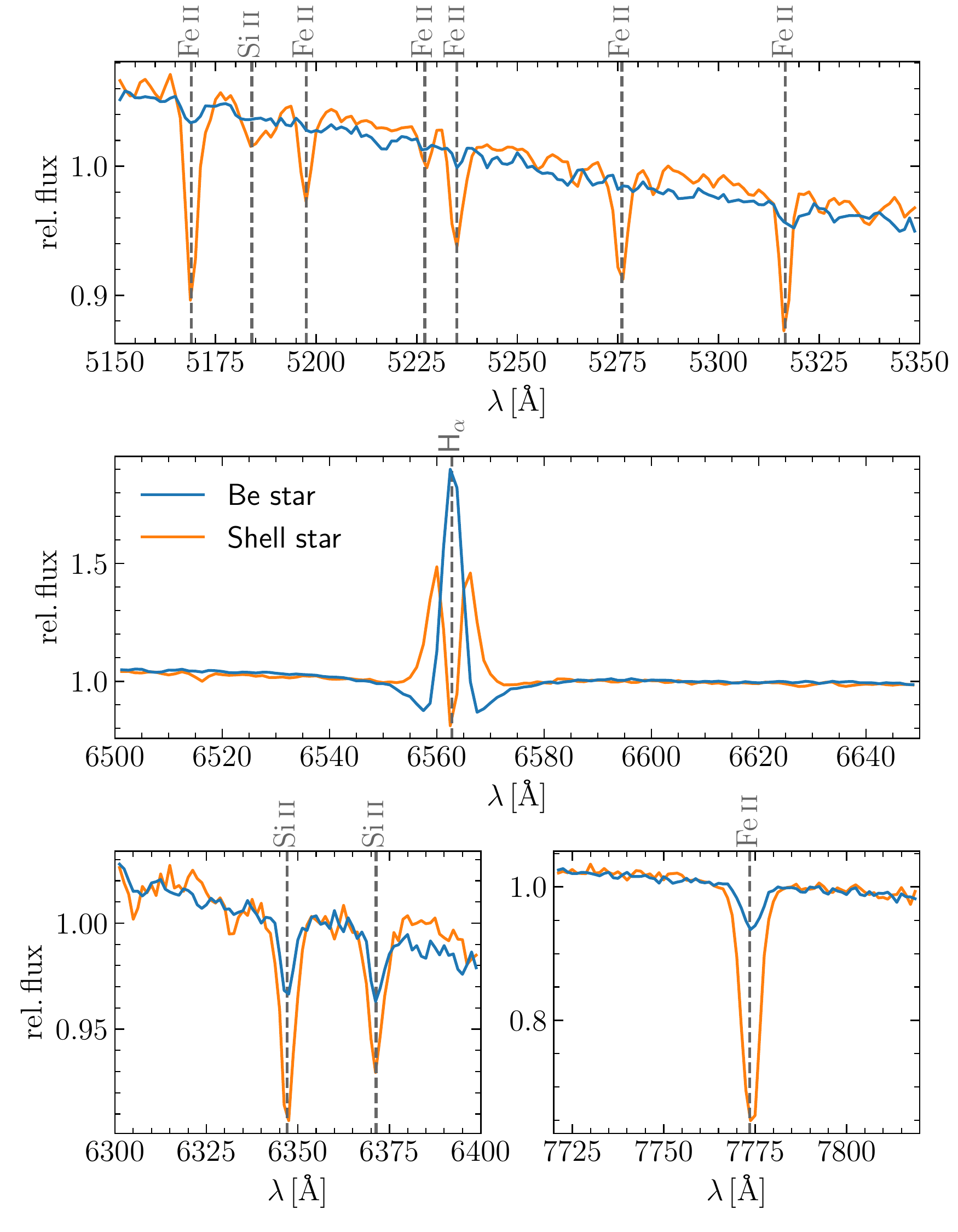}
    \caption{Comparison of a ``standard'' Be star spectrum (star number 92) and a prototypical shell star spectrum (star number 5\,000\,003) from our sample. Vertical dashed lines represent absorption features used to identify the shell stars.}
    \label{fig:example_spectra}
\end{figure}

Visual inspection of some of the Be-star spectra revealed that they could roughly be divided into two classes, as illustrated in Fig.~\ref{fig:example_spectra}. Besides the ``normal'' Be-stars spectra, which are characterized by H$\alpha$ emission, broad Paschen lines, and few \ion{He}{i} lines, we identified a group of spectra that showed a pronounced double peak in the H$\alpha$ line profile, a series of very narrow Paschen lines extending to much higher orders than for the ``normal'' Be-star spectra, and a number of narrow \ion{Fe}{ii} and \ion{Si}{ii} absorption lines. These features are characteristic for shell stars, Be stars observed (almost) equator-on such that that the star is seen through the decretion disk \citep[e.g.,][]{rivinius06}.

We noted that the strengths of the various shell features (double-peaked H$\alpha$ line, high-order Paschen lines, \ion{Fe}{ii} and \ion{Si}{ii} lines) varied from star to star. This gradual transition can be understood in two ways. First, for decreasing inclinations, the fraction of the photosphere that is observed through the disk decreases. Second, the amount of matter in the disks is time-dependent. In order to obtain an unbiased estimate of the number of shell stars in our sample, we performed a cross-correlation of all spectroscopic Be stars against the two prototypical spectra shown in Fig.~\ref{fig:example_spectra}. Then, we compared the strengths of the two cross correlation signals using the $r$ parameter as defined by \citet{tonry79} and classified as shell stars all stars for which $r_{\rm shell}>5$ and $r_{\rm shell}/r_{\rm Be}>1.5$. The 47/202 stars classified this way are highlighted as green diamonds in Fig.~\ref{fig:halpha_cmd}.  It is evident that compared to normal Be stars at the same magnitude, shell stars show higher H$\alpha$ equivalent widths. Furthermore, the shell stars appear fainter in $m_{\rm F438W}$ than the normal Be stars on average. These trends appear to be in agreement with the expectations from absorption in a disk.

As mentioned previously, in our sample of 202 spectroscopic Be stars, we classified 47 as shell stars, giving a shell-star fraction of $23$\%.  This fraction can be used to estimate the opening angle of the disk, under the assumption of a random distribution of spin axes \citep[as found by][for another young massive cluster, NGC~330]{hummel99}, resulting in a half-opening angle of $\phi=13$~degrees. The same value was derived by \citet{Hanuschik96}, who used a sample of 114 Galactic Be (including 26 Shell stars) stars to estimate the shell-star fraction and obtained 22.8\%. The good agreement suggests that there is not a strong effect of environment or metallicity on the Be star disk opening angle (at least in the comparison between the Milky Way field and NGC~1850 in the LMC).

We note that, as shown by \citet{cyr15}, the inferred opening angles depend on the diagnostic applied and the disk geometry adopted. The value derived for NGC~1850 is still within the confidence intervals provided by \citet{cyr15}.

\subsection{Implications for age spreads within clusters}
\label{sec:implications}

\begin{figure}
    \centering
    \includegraphics[width=.95\linewidth]{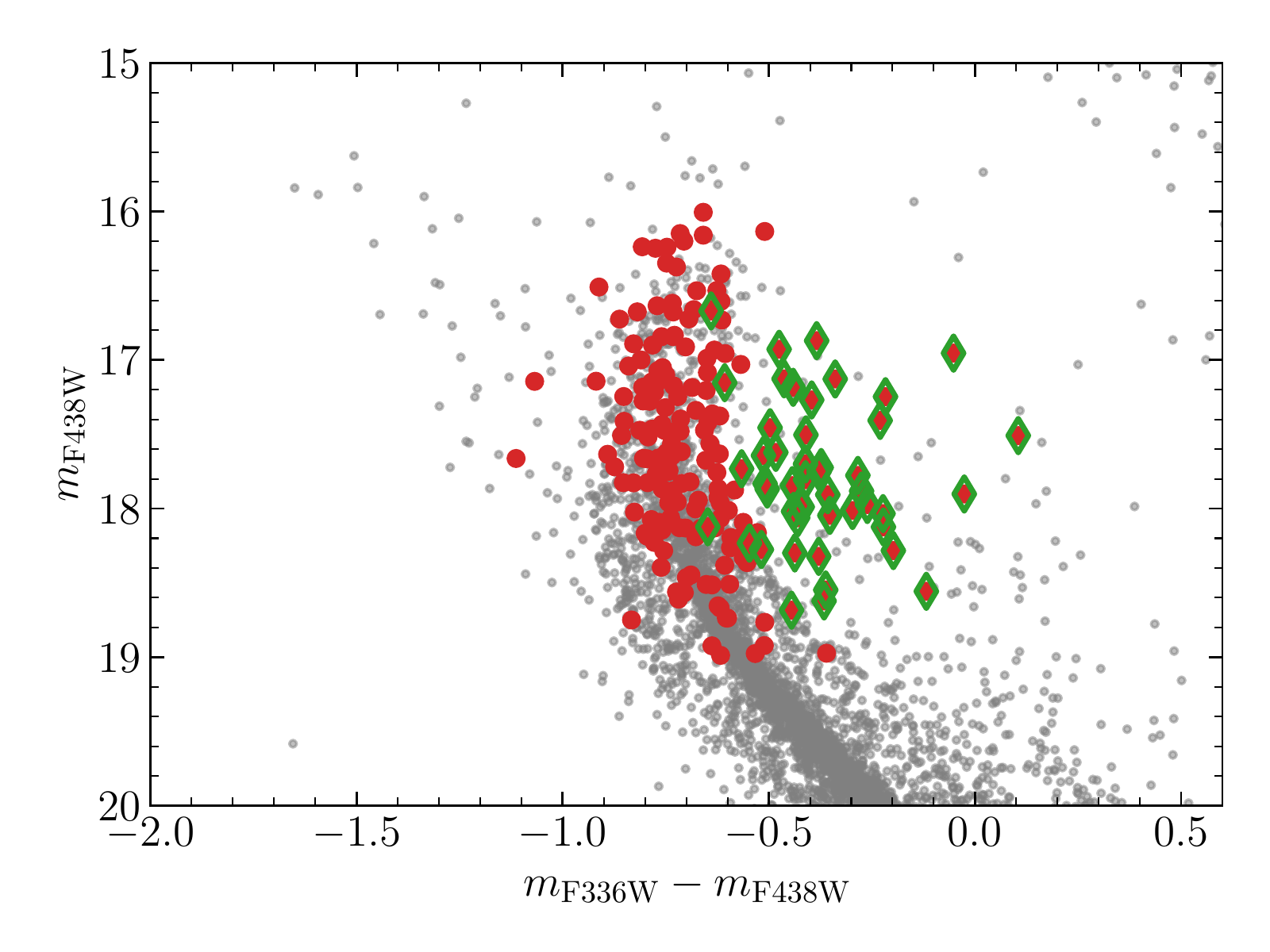}
    \caption{HST based colour magnitude diagram of NGC 1850, with likely members shown as grey points, and Be and shell stars highlighted as red points and green diamonds, respectively.  Note that Be stars are concentrated towards the main sequence turn-off (MSTO) and the shell stars lie to the red of the MSTO.}
    \label{fig:cmd_be_shell}
\end{figure}

The location of the shell stars in a ($m_{\rm F336W} - m_{\rm F438W}$, $m_{\rm F438W}$) CMD is shown in Fig.~\ref{fig:cmd_be_shell}. It can be seen that the shell stars lie redder and fainter than the nominal MSTO within NGC~1850.  As discussed above, this is likely due to the extinction of the host star when seen nearly equator on. Given the position of the shell stars within the CMD of NGC~1850, it appears that the effect of self-extinction of Be stars can be larger than the change in colour and magnitude due to the effects of stellar rotation on the stars themselves.

To date, stellar models that include rotation do not account for the influence of the decretion disks for rapidly rotating stars. Furthermore, disks are a transient phenomenon and build up and dissipate on timescales comparable to the stellar activity cycles. As a consequence, Be stars have no fixed positions in a CMD, but instead follow loops with outlines that are mainly determined by their inclinations \citep{dewit06,haubois12,rimulo18}. Hence, the current positions of Be or shell stars in the observed CMDs are not expected to be matched by models, even those including rotational effects.  The expectation is therefore that the width of the MSTO will be larger than predicted by models that include a wide range of rotational velocities. 

\citet{correnti17} have studied NGC~1850, comparing the observed HST CMDs with isochrones that include stellar rotation.  These authors conclude that the MSTO width is larger than can be explained through stellar rotation alone, hence that an age spread of $\sim35$~Myr must be present within the cluster.  However, the flux excesses from decretion disks, and in particular the presence of significant numbers of shell stars within the cluster, which are not represented in the stellar models, calls into question the need for a major age spread within the cluster \footnote{We point the interested reader to the review of \citet{BL18} for a discussion of the evidence against significant age spreads within massive clusters.}.

The same shell star phenomenon is also expected to happen for lower mass stars (down to $\sim1.5$~\msun), although potentially at lower rates.  It would be more difficult to find these lower mass stars with disks, because, due to their lower temperatures, they will not ionise their disks, i.e., the disks will not display emission lines.  However, such disks can still produce shell absorption lines. In addition, the disks should be detectable through their mid-IR colours \citep[e.g.,][]{kenyon95}, a promising avenue for future work when higher spatial resolution mid-IR observatories come online.

\subsection{Rotation rates of Be and shell stars}
\label{sec:be_rotation}

\begin{figure}
    \centering
    \includegraphics[width=.95\linewidth]{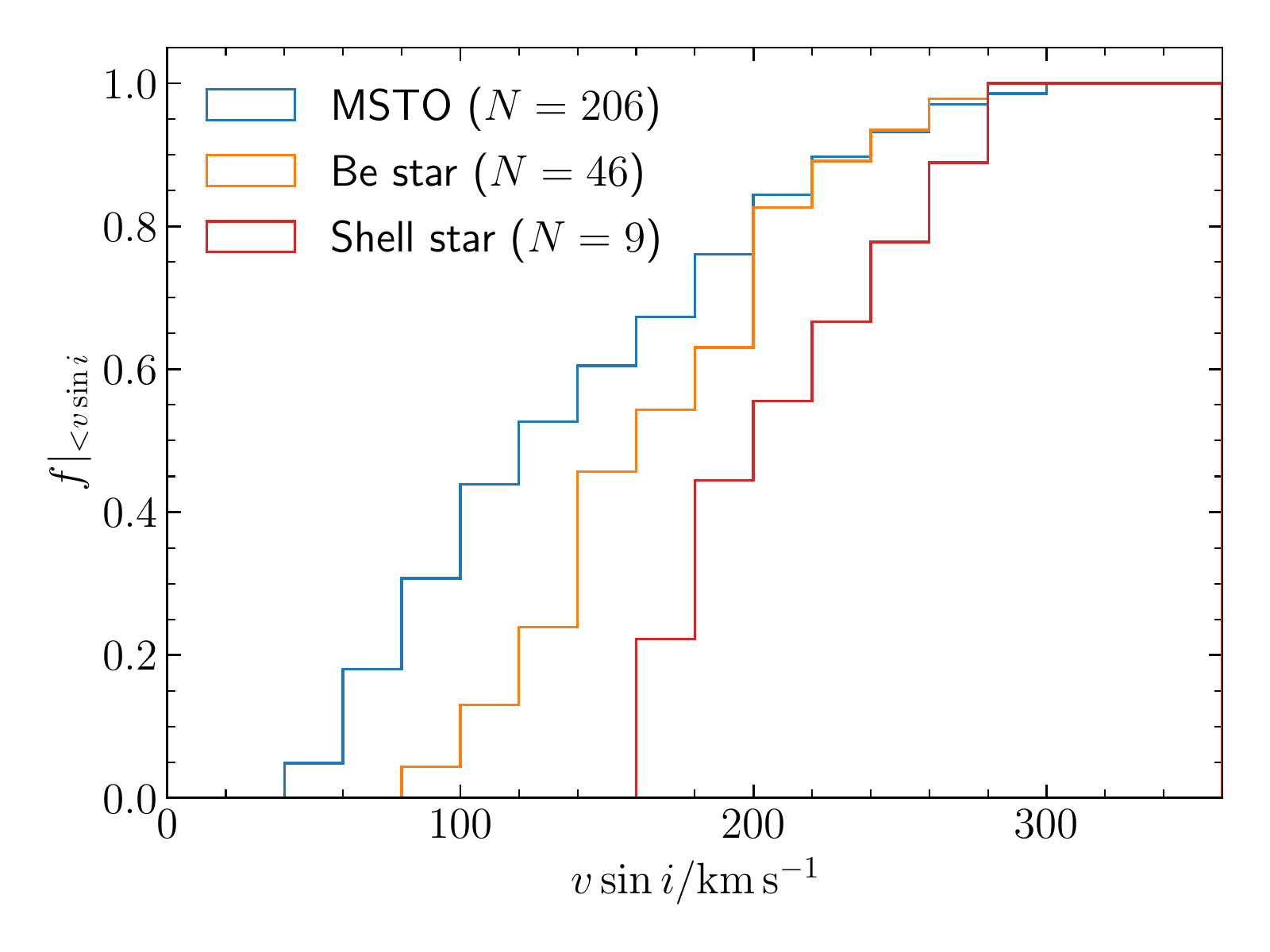}
    \caption{Cumulative distributions of the \vsini{} measurements obtained for Be stars (orange), shell stars (red), and normal MSTO stars (blue).  For each group, the number of stars with valid \vsini{} measurements is provided in the legend.}
    \label{fig:vsini_be}
\end{figure}

In Fig.~\ref{fig:vsini_be}, we show the \vsini{} measurements obtained for the Be and shell stars in our sample and compare them to the results obtained for normal MSTO stars. As explained in Sec.~\ref{sec:rotation:individual}, we discarded the \textsc{Spexxy} results for Be and shell stars, given the lack of any adequate models in the \citet{ferre} spectral library. To enable a better comparison to the normal MSTO stars, we therefore show in Fig.~\ref{fig:vsini_be} only results based on individual line fits. Fig.~\ref{fig:vsini_be} shows that on average, Be stars rotate faster than normal main sequence stars, as expected. The tail towards low \vsini{} values observed for the Be stars is likely produced by stars observed at low inclination $i$. Given that Be stars observed at low inclination should not show shell features, it is reassuring that no shell stars with \vsini{}$<160\,{\rm km\,s^{-1}}$ are observed.

It is interesting to note that the \vsini{} distributions of Be and shell stars shown in Fig.~\ref{fig:vsini_be} do not extend to higher values compared to normal MSTO stars. This would have been expected if all near-critically rotating MSTO stars appear as Be stars. Our results support the idea that fast rotation alone is not a sufficient condition for a star to show a decretion disk, either because the disk is a transient feature or because additional mechanisms are needed to trigger disk formation \citep[e.g.,][]{baade2020}. Furthermore, a comparison of the histograms shown in Fig.~\ref{fig:vsini_be} to the critical rotation velocity predicted for the MSTO of NGC~1850, $V_{\rm crit}\approx400\,{\rm km\,s^{-1}}$ (cf. Sec.~\ref{sec:rotation:models}) suggests that the Be stars are not rotating close to the expected break-up velocity. Recently, \citet{dufton22} found an average ratio of $V/V_{\rm crit}=0.68$ for a sample of (more massive) Be stars in the 30~Doradus region, while \citet{rivinius2013} found an average value of $\sim0.7$ for the Galactic field when compiling various literature studies. While the distributions shown in Fig.~\ref{fig:vsini_be} also extend to $V/V_{\rm crit}\sim0.0.7$, they suggest a somewhat lower average ratio in NGC~1850.

\section{Conclusions}
\label{sec:conclusions}

We have presented an analysis of multi-epoch MUSE spectroscopy of the stellar populations within the $\sim$100~Myr LMC star cluster, NGC~1850.  Our full sample consists of more than 4000 stars, and each star has been observed for a number of typically 16 epochs. For the present work, we have combined the single-epoch spectra on a star-by-star basis and, following cuts based on S/N and cluster membership probabilities, ended up with a sample of \nstars{} stars with MUSE spectra that we analysed in order to understand the distribution of stellar rotation across the stellar population of NGC~1850. The set of combined MUSE spectra, along with the HST photometry, is made publicly available\footnote{Link to VizieR}.

The main results can be summarised as follows:

\begin{itemize}
    \item There is a clear correlation between the colour of stars on the MSTO and their observed \vsini{} value.  This confirms previous suggestions that the extended MSTO phenomenon is driven by the stellar rotation distribution \citep{bastian09}.
    \item The two branches of the split main sequence have markedly different \vsini{} distributions, with the blue arm made up primarily of slow rotators while the red arm consists mainly of rapid rotators. Similar bimodal distributions have also been found for B stars in the Galactic field \citep[e.g.,][]{sun21} or the 30~Doradus region of the LMC \citep{dufton13}. Our result confirms previous results found for other clusters, based on much smaller stellar samples and confirms predictions of stellar models that include rotation \citep[e.g.,][]{dantona15}.
    \item Over a relatively large range in magnitude ($17<m_{\rm F438W}<20$), we find that the fast rotating stars have a median \vsini{} of $\sim200\,{\rm km\,s^{-1}}$ and that the \vsini{} distribution is barely populated beyond $\sim300\,{\rm km\,s^{-1}}$. Comparison to stellar models suggests that not many main sequence stars are rotating close to their break up velocities. However, the evolution of the latter for stars approaching the MSTO suggests that a large fraction of fast rotators could be critically rotating at the end of their main sequence lifetimes. Interestingly though, the MSTO of NGC~1850 shows a lack of stars with \vsini{} values close to the predicted critical value of $400\,{\rm km\,s^{-1}}$.
    \item Similar to previous analyses \citep[e.g.][]{milone18_be}, we find that the Be star fraction is a strong function of magnitude, increasing towards the MSTO, and going to zero on the nominal main sequence. In light of the predicted drop in critical velocity depicted in Fig.~\ref{fig:syclist}, this appears in qualitative agreement with Be model predictions based on single star evolution \citep[see also][]{hastings20}.  However, under certain assumptions, a similar trend can also be explained in a binary driven scenario (see the discussion in \citealt{hastings21}). A promising avenue for future research is the comparison of rotation velocities of main sequence stars across LMC clusters of different ages, as they provide snapshots of very similar stars (in terms of mass, metallicity, and environment) at different timestamps of their main sequence evolution. 
    \item Within the Be star population, we find $23$\% of the stars to be "shell stars", i.e., Be stars that are seen nearly equator-on.  By assuming an isotropic distribution of the Be stars' spin axes, we can translate this fraction to the Be star decretion disk opening angle, for which we find $\sim15$~degrees (half opening angle), in agreement to what has been found for Galactic Be stars.
    \item  The shell stars are found almost exclusively on the red side of the MSTO, suggesting that they are self-extincted by their own disks. The photometric shifts are significantly larger than those predicted for ``normal'' Be stars owing to the growth and waning of their disks with time. Since decretion disks are not included in stellar models (even those that include rotation), the spread on the MSTO will never be matched by stellar models alone. This may result in the spurious inference of age spreads within clusters that host Be stars. 
    \item This phenomenon is also likely to be present in older clusters (with ages up to $\sim1.5-2.0$~Gyr), although the stars with disks will be more difficult to identify (since they will not be hot enough to ionise their disks to be observed as late type A or F 'e' (emission) stars).  Such stars should be identifiable, however, through their mid-IR colours, via shell-absorption lines or filled-in Balmer lines \citep{slettebak82}. Studying LMC clusters of different ages again appears as a promising avenue for future research, in particular given that the Balmer lines are much easier accessible in clusters other than NGC~1850, where the strong nebulosity complicates matters.

\end{itemize}

\section*{Acknowledgements}

We thank the anonymous referee for a careful reading of our manuscript and for their helpful comments.
SK acknowledges funding from UKRI in the form of a Future Leaders Fellowship (grant no. MR/T022868/1).
SS acknowledges funding from STFC under the grant no. R276234.
NB gratefully acknowledges funding from the ERC. This study was supported by the Klaus Tschira Foundation.
CU acknowledges the support of the Swedish Research Council, Vetenskapsr{\aa}det.
SE and CG acknowledge support through the STAREX grant from the ERC Horizon 2020 research and innovation programme (grant agreement no. 833925).
\section*{Data Availability}

The HST photometry and the combined MUSE spectra are available on VizieR. All other data underlying this article will be shared on reasonable request to the corresponding author.



\bibliographystyle{mnras}
\bibliography{ngc1850_rotation} 

\begin{thebibliography}{}
\makeatletter
\relax
\def\mn@urlcharsother{\let\do\@makeother \do\$\do\&\do\#\do\^\do\_\do\%\do\~}
\def\mn@doi{\begingroup\mn@urlcharsother \@ifnextchar [ {\mn@doi@}
  {\mn@doi@[]}}
\def\mn@doi@[#1]#2{\def\@tempa{#1}\ifx\@tempa\@empty \href
  {http://dx.doi.org/#2} {doi:#2}\else \href {http://dx.doi.org/#2} {#1}\fi
  \endgroup}
\def\mn@eprint#1#2{\mn@eprint@#1:#2::\@nil}
\def\mn@eprint@arXiv#1{\href {http://arxiv.org/abs/#1} {{\tt arXiv:#1}}}
\def\mn@eprint@dblp#1{\href {http://dblp.uni-trier.de/rec/bibtex/#1.xml}
  {dblp:#1}}
\def\mn@eprint@#1:#2:#3:#4\@nil{\def\@tempa {#1}\def\@tempb {#2}\def\@tempc
  {#3}\ifx \@tempc \@empty \let \@tempc \@tempb \let \@tempb \@tempa \fi \ifx
  \@tempb \@empty \def\@tempb {arXiv}\fi \@ifundefined
  {mn@eprint@\@tempb}{\@tempb:\@tempc}{\expandafter \expandafter \csname
  mn@eprint@\@tempb\endcsname \expandafter{\@tempc}}}

\bibitem[\protect\citeauthoryear{{Abt} \& {Boonyarak}}{{Abt} \&
  {Boonyarak}}{2004}]{abt04}
{Abt} H.~A.,  {Boonyarak} C.,  2004, \mn@doi [\apj] {10.1086/423795}, \href
  {https://ui.adsabs.harvard.edu/abs/2004ApJ...616..562A} {616, 562}

\bibitem[\protect\citeauthoryear{{Abt}, {Levato}  \& {Grosso}}{{Abt}
  et~al.}{2002}]{abt02}
{Abt} H.~A.,  {Levato} H.,   {Grosso} M.,  2002, \mn@doi [\apj]
  {10.1086/340590}, \href
  {https://ui.adsabs.harvard.edu/abs/2002ApJ...573..359A} {573, 359}

\bibitem[\protect\citeauthoryear{{Allende Prieto}, {Koesterke}, {Hubeny},
  {Bautista}, {Barklem}  \& {Nahar}}{{Allende Prieto} et~al.}{2018}]{ferre}
{Allende Prieto} C.,  {Koesterke} L.,  {Hubeny} I.,  {Bautista} M.~A.,
  {Barklem} P.~S.,   {Nahar} S.~N.,  2018, \mn@doi [\aap]
  {10.1051/0004-6361/201732484}, \href
  {https://ui.adsabs.harvard.edu/abs/2018A&A...618A..25A} {618, A25}

\bibitem[\protect\citeauthoryear{{Arentsen} et~al.,}{{Arentsen}
  et~al.}{2019}]{arentsen2019}
{Arentsen} A.,  et~al., 2019, \mn@doi [\aap] {10.1051/0004-6361/201834273},
  \href {https://ui.adsabs.harvard.edu/abs/2019A&A...627A.138A} {627, A138}

\bibitem[\protect\citeauthoryear{{Baade} \& {Rivinius}}{{Baade} \&
  {Rivinius}}{2020}]{baade2020}
{Baade} D.,  {Rivinius} T.,  2020, in {Neiner} C.,  {Weiss} W.~W.,  {Baade} D.,
   {Griffin} R.~E.,  {Lovekin} C.~C.,   {Moffat} A.~F.~J.,  eds, Stars and
  their Variability Observed from Space. pp 35--38

\bibitem[\protect\citeauthoryear{{Bacon} et~al.,}{{Bacon}
  et~al.}{2010}]{2010SPIE.7735E..08B}
{Bacon} R.,  et~al., 2010, in Ground-based and Airborne Instrumentation for
  Astronomy III. p. 773508, \mn@doi{10.1117/12.856027}

\bibitem[\protect\citeauthoryear{{Bailey} \& {Landstreet}}{{Bailey} \&
  {Landstreet}}{2013}]{bailey13}
{Bailey} J.~D.,  {Landstreet} J.~D.,  2013, \mn@doi [\aap]
  {10.1051/0004-6361/201220671}, \href
  {https://ui.adsabs.harvard.edu/abs/2013A&A...551A..30B} {551, A30}

\bibitem[\protect\citeauthoryear{{Bastian} \& {Lardo}}{{Bastian} \&
  {Lardo}}{2018}]{BL18}
{Bastian} N.,  {Lardo} C.,  2018, \mn@doi [\araa]
  {10.1146/annurev-astro-081817-051839}, \href
  {https://ui.adsabs.harvard.edu/abs/2018ARA&A..56...83B} {56, 83}

\bibitem[\protect\citeauthoryear{{Bastian} \& {de Mink}}{{Bastian} \& {de
  Mink}}{2009}]{bastian09}
{Bastian} N.,  {de Mink} S.~E.,  2009, \mn@doi [\mnras]
  {10.1111/j.1745-3933.2009.00696.x}, \href
  {https://ui.adsabs.harvard.edu/abs/2009MNRAS.398L..11B} {398, L11}

\bibitem[\protect\citeauthoryear{{Bastian} et~al.,}{{Bastian}
  et~al.}{2017}]{bastian17}
{Bastian} N.,  et~al., 2017, \mn@doi [\mnras] {10.1093/mnras/stw3042}, \href
  {https://ui.adsabs.harvard.edu/abs/2017MNRAS.465.4795B} {465, 4795}

\bibitem[\protect\citeauthoryear{{Bastian}, {Kamann}, {Cabrera-Ziri}, {Georgy},
  {Ekstr{\"o}m}, {Charbonnel}, {de Juan Ovelar}  \& {Usher}}{{Bastian}
  et~al.}{2018}]{bastian18_n2818}
{Bastian} N.,  {Kamann} S.,  {Cabrera-Ziri} I.,  {Georgy} C.,  {Ekstr{\"o}m}
  S.,  {Charbonnel} C.,  {de Juan Ovelar} M.,   {Usher} C.,  2018, \mn@doi
  [\mnras] {10.1093/mnras/sty2100}, \href
  {https://ui.adsabs.harvard.edu/abs/2018MNRAS.480.3739B} {480, 3739}

\bibitem[\protect\citeauthoryear{{Bastian}, {Kamann}, {Amard}, {Charbonnel},
  {Haemmerl{\'e}}  \& {Matt}}{{Bastian} et~al.}{2020}]{bastian20}
{Bastian} N.,  {Kamann} S.,  {Amard} L.,  {Charbonnel} C.,  {Haemmerl{\'e}} L.,
    {Matt} S.~P.,  2020, \mn@doi [\mnras] {10.1093/mnras/staa1332}, \href
  {https://ui.adsabs.harvard.edu/abs/2020MNRAS.495.1978B} {495, 1978}

\bibitem[\protect\citeauthoryear{{Bertin}, {Mellier}, {Radovich}, {Missonnier},
  {Didelon}  \& {Morin}}{{Bertin} et~al.}{2002}]{2002ASPC..281..228B}
{Bertin} E.,  {Mellier} Y.,  {Radovich} M.,  {Missonnier} G.,  {Didelon} P.,
  {Morin} B.,  2002, in {Bohlender} D.~A.,  {Durand} D.,   {Handley} T.~H.,
  eds,  Astronomical Society of the Pacific Conference Series Vol. 281,
  Astronomical Data Analysis Software and Systems XI. p.~228

\bibitem[\protect\citeauthoryear{{Bodensteiner} et~al.,}{{Bodensteiner}
  et~al.}{2020a}]{bodensteiner20b}
{Bodensteiner} J.,  et~al., 2020a, \mn@doi [\aap]
  {10.1051/0004-6361/201936743}, \href
  {https://ui.adsabs.harvard.edu/abs/2020A&A...634A..51B} {634, A51}

\bibitem[\protect\citeauthoryear{{Bodensteiner}, {Shenar}  \&
  {Sana}}{{Bodensteiner} et~al.}{2020b}]{Bodensteiner20}
{Bodensteiner} J.,  {Shenar} T.,   {Sana} H.,  2020b, \mn@doi [\aap]
  {10.1051/0004-6361/202037640}, \href
  {https://ui.adsabs.harvard.edu/abs/2020A&A...641A..42B} {641, A42}

\bibitem[\protect\citeauthoryear{{Boubert} \& {Evans}}{{Boubert} \&
  {Evans}}{2018}]{boubert2018}
{Boubert} D.,  {Evans} N.~W.,  2018, \mn@doi [\mnras] {10.1093/mnras/sty980},
  \href {https://ui.adsabs.harvard.edu/abs/2018MNRAS.477.5261B} {477, 5261}

\bibitem[\protect\citeauthoryear{{Cadelano}, {Dalessandro}, {Salaris},
  {Bastian}, {Mucciarelli}, {Saracino}, {Martocchia}  \&
  {Cabrera-Ziri}}{{Cadelano} et~al.}{2022}]{cadelano2022}
{Cadelano} M.,  {Dalessandro} E.,  {Salaris} M.,  {Bastian} N.,  {Mucciarelli}
  A.,  {Saracino} S.,  {Martocchia} S.,   {Cabrera-Ziri} I.,  2022, \mn@doi
  [\apjl] {10.3847/2041-8213/ac424a}, \href
  {https://ui.adsabs.harvard.edu/abs/2022ApJ...924L...2C} {924, L2}

\bibitem[\protect\citeauthoryear{{Cordoni}, {Milone}, {Marino}, {Di
  Criscienzo}, {D'Antona}, {Dotter}, {Lagioia}  \& {Tailo}}{{Cordoni}
  et~al.}{2018}]{cordoni2018}
{Cordoni} G.,  {Milone} A.~P.,  {Marino} A.~F.,  {Di Criscienzo} M.,
  {D'Antona} F.,  {Dotter} A.,  {Lagioia} E.~P.,   {Tailo} M.,  2018, \mn@doi
  [\apj] {10.3847/1538-4357/aaedc1}, \href
  {https://ui.adsabs.harvard.edu/abs/2018ApJ...869..139C} {869, 139}

\bibitem[\protect\citeauthoryear{{Correnti}, {Goudfrooij}, {Bellini}, {Kalirai}
   \& {Puzia}}{{Correnti} et~al.}{2017}]{correnti17}
{Correnti} M.,  {Goudfrooij} P.,  {Bellini} A.,  {Kalirai} J.~S.,   {Puzia}
  T.~H.,  2017, \mn@doi [\mnras] {10.1093/mnras/stx010}, \href
  {https://ui.adsabs.harvard.edu/abs/2017MNRAS.467.3628C} {467, 3628}

\bibitem[\protect\citeauthoryear{{Cyr}, {Jones}  \& {Tycner}}{{Cyr}
  et~al.}{2015}]{cyr15}
{Cyr} R.~P.,  {Jones} C.~E.,   {Tycner} C.,  2015, \mn@doi [\apj]
  {10.1088/0004-637X/799/1/33}, \href
  {https://ui.adsabs.harvard.edu/abs/2015ApJ...799...33C} {799, 33}

\bibitem[\protect\citeauthoryear{{Czesla}, {Schr{\"o}ter}, {Schneider},
  {Huber}, {Pfeifer}, {Andreasen}  \& {Zechmeister}}{{Czesla}
  et~al.}{2019}]{pya}
{Czesla} S.,  {Schr{\"o}ter} S.,  {Schneider} C.~P.,  {Huber} K.~F.,  {Pfeifer}
  F.,  {Andreasen} D.~T.,   {Zechmeister} M.,  2019, {PyA: Python
  astronomy-related packages} (\mn@eprint {ascl} {1906.010})

\bibitem[\protect\citeauthoryear{{D'Antona}, {Di Criscienzo}, {Decressin},
  {Milone}, {Vesperini}  \& {Ventura}}{{D'Antona} et~al.}{2015}]{dantona15}
{D'Antona} F.,  {Di Criscienzo} M.,  {Decressin} T.,  {Milone} A.~P.,
  {Vesperini} E.,   {Ventura} P.,  2015, \mn@doi [\mnras]
  {10.1093/mnras/stv1794}, \href
  {https://ui.adsabs.harvard.edu/abs/2015MNRAS.453.2637D} {453, 2637}

\bibitem[\protect\citeauthoryear{{Dolphin}}{{Dolphin}}{2000}]{Dolphin00}
{Dolphin} A.~E.,  2000, \mn@doi [\pasp] {10.1086/316630}, \href
  {https://ui.adsabs.harvard.edu/abs/2000PASP..112.1383D} {112, 1383}

\bibitem[\protect\citeauthoryear{{Dolphin}}{{Dolphin}}{2016}]{Dolphin16}
{Dolphin} A.,  2016, {DOLPHOT: Stellar photometry} (\mn@eprint {ascl}
  {1608.013})

\bibitem[\protect\citeauthoryear{{Dotter}}{{Dotter}}{2016}]{mist0}
{Dotter} A.,  2016, \mn@doi [\apjs] {10.3847/0067-0049/222/1/8}, \href
  {https://ui.adsabs.harvard.edu/abs/2016ApJS..222....8D} {222, 8}

\bibitem[\protect\citeauthoryear{{Dufton} et~al.,}{{Dufton}
  et~al.}{2013}]{dufton13}
{Dufton} P.~L.,  et~al., 2013, \mn@doi [\aap] {10.1051/0004-6361/201220273},
  \href {https://ui.adsabs.harvard.edu/abs/2013A&A...550A.109D} {550, A109}

\bibitem[\protect\citeauthoryear{{Dufton}, {Lennon}, {Villase{\~n}or},
  {Howarth}, {Evans}, {de Mink}, {Sana}  \& {Taylor}}{{Dufton}
  et~al.}{2022}]{dufton22}
{Dufton} P.~L.,  {Lennon} D.~J.,  {Villase{\~n}or} J.~I.,  {Howarth} I.~D.,
  {Evans} C.~J.,  {de Mink} S.~E.,  {Sana} H.,   {Taylor} W.~D.,  2022, \mn@doi
  [\mnras] {10.1093/mnras/stac630}, \href
  {https://ui.adsabs.harvard.edu/abs/2022MNRAS.512.3331D} {512, 3331}

\bibitem[\protect\citeauthoryear{{Dupree} et~al.,}{{Dupree}
  et~al.}{2017}]{dupree17}
{Dupree} A.~K.,  et~al., 2017, \mn@doi [\apjl] {10.3847/2041-8213/aa85dd},
  \href {https://ui.adsabs.harvard.edu/abs/2017ApJ...846L...1D} {846, L1}

\bibitem[\protect\citeauthoryear{{Ekstr{\"o}m}, {Meynet}, {Maeder}  \&
  {Barblan}}{{Ekstr{\"o}m} et~al.}{2008}]{ekstrom08}
{Ekstr{\"o}m} S.,  {Meynet} G.,  {Maeder} A.,   {Barblan} F.,  2008, \mn@doi
  [\aap] {10.1051/0004-6361:20078095}, \href
  {https://ui.adsabs.harvard.edu/abs/2008A&A...478..467E} {478, 467}

\bibitem[\protect\citeauthoryear{{El-Badry} \& {Burdge}}{{El-Badry} \&
  {Burdge}}{2022}]{elbadry22}
{El-Badry} K.,  {Burdge} K.~B.,  2022, \mn@doi [\mnras]
  {10.1093/mnrasl/slab135}, \href
  {https://ui.adsabs.harvard.edu/abs/2022MNRAS.511L..24E} {511, 24}

\bibitem[\protect\citeauthoryear{{Espinosa Lara} \& {Rieutord}}{{Espinosa Lara}
  \& {Rieutord}}{2011}]{espinosa_lara2011}
{Espinosa Lara} F.,  {Rieutord} M.,  2011, \mn@doi [\aap]
  {10.1051/0004-6361/201117252}, \href
  {https://ui.adsabs.harvard.edu/abs/2011A&A...533A..43E} {533, A43}

\bibitem[\protect\citeauthoryear{{Fabregat} \& {Torrej{\'o}n}}{{Fabregat} \&
  {Torrej{\'o}n}}{2000}]{fabregat2000}
{Fabregat} J.,  {Torrej{\'o}n} J.~M.,  2000, \aap, \href
  {https://ui.adsabs.harvard.edu/abs/2000A&A...357..451F} {357, 451}

\bibitem[\protect\citeauthoryear{{Feast}}{{Feast}}{1972}]{feast72}
{Feast} M.~W.,  1972, \mn@doi [\mnras] {10.1093/mnras/159.2.113}, \href
  {https://ui.adsabs.harvard.edu/abs/1972MNRAS.159..113F} {159, 113}

\bibitem[\protect\citeauthoryear{{Foreman-Mackey}, {Hogg}, {Lang}  \&
  {Goodman}}{{Foreman-Mackey} et~al.}{2013}]{emcee}
{Foreman-Mackey} D.,  {Hogg} D.~W.,  {Lang} D.,   {Goodman} J.,  2013, \mn@doi
  [\pasp] {10.1086/670067}, \href
  {https://ui.adsabs.harvard.edu/abs/2013PASP..125..306F} {125, 306}

\bibitem[\protect\citeauthoryear{{Georgy}, {Ekstr{\"o}m}, {Granada}, {Meynet},
  {Mowlavi}, {Eggenberger}  \& {Maeder}}{{Georgy} et~al.}{2013}]{georgy2013}
{Georgy} C.,  {Ekstr{\"o}m} S.,  {Granada} A.,  {Meynet} G.,  {Mowlavi} N.,
  {Eggenberger} P.,   {Maeder} A.,  2013, \mn@doi [\aap]
  {10.1051/0004-6361/201220558}, \href
  {https://ui.adsabs.harvard.edu/abs/2013A&A...553A..24G} {553, A24}

\bibitem[\protect\citeauthoryear{{Giesers} et~al.,}{{Giesers}
  et~al.}{2019}]{giesers2019}
{Giesers} B.,  et~al., 2019, \mn@doi [\aap] {10.1051/0004-6361/201936203},
  \href {https://ui.adsabs.harvard.edu/abs/2019A&A...632A...3G} {632, A3}

\bibitem[\protect\citeauthoryear{{Gonneau} et~al.,}{{Gonneau}
  et~al.}{2020}]{gonneau2020}
{Gonneau} A.,  et~al., 2020, \mn@doi [\aap] {10.1051/0004-6361/201936825},
  \href {https://ui.adsabs.harvard.edu/abs/2020A&A...634A.133G} {634, A133}

\bibitem[\protect\citeauthoryear{{Goodman} \& {Weare}}{{Goodman} \&
  {Weare}}{2010}]{goodman2010}
{Goodman} J.,  {Weare} J.,  2010, \mn@doi [Communications in Applied
  Mathematics and Computational Science] {10.2140/camcos.2010.5.65}, \href
  {https://ui.adsabs.harvard.edu/abs/2010CAMCS...5...65G} {5, 65}

\bibitem[\protect\citeauthoryear{{Gossage} et~al.,}{{Gossage}
  et~al.}{2019}]{Gossage19}
{Gossage} S.,  et~al., 2019, \mn@doi [\apj] {10.3847/1538-4357/ab5717}, \href
  {https://ui.adsabs.harvard.edu/abs/2019ApJ...887..199G} {887, 199}

\bibitem[\protect\citeauthoryear{{Granada}, {Ekstr{\"o}m}, {Georgy},
  {Krti{\v{c}}ka}, {Owocki}, {Meynet}  \& {Maeder}}{{Granada}
  et~al.}{2013}]{granada2013}
{Granada} A.,  {Ekstr{\"o}m} S.,  {Georgy} C.,  {Krti{\v{c}}ka} J.,  {Owocki}
  S.,  {Meynet} G.,   {Maeder} A.,  2013, \mn@doi [\aap]
  {10.1051/0004-6361/201220559}, \href
  {https://ui.adsabs.harvard.edu/abs/2013A&A...553A..25G} {553, A25}

\bibitem[\protect\citeauthoryear{{Gratton}, {Bragaglia}, {Carretta}, {D'Orazi},
  {Lucatello}  \& {Sollima}}{{Gratton} et~al.}{2019}]{gratton_review}
{Gratton} R.,  {Bragaglia} A.,  {Carretta} E.,  {D'Orazi} V.,  {Lucatello} S.,
   {Sollima} A.,  2019, \mn@doi [\aapr] {10.1007/s00159-019-0119-3}, \href
  {https://ui.adsabs.harvard.edu/abs/2019A&ARv..27....8G} {27, 8}

\bibitem[\protect\citeauthoryear{{Gray}}{{Gray}}{2008}]{gray2008}
{Gray} D.~F.,  2008, {The Observation and Analysis of Stellar Photospheres}

\bibitem[\protect\citeauthoryear{{Grebel}, {Richter}  \& {de Boer}}{{Grebel}
  et~al.}{1992}]{grebel92}
{Grebel} E.~K.,  {Richter} T.,   {de Boer} K.~S.,  1992, \aap, \href
  {https://ui.adsabs.harvard.edu/abs/1992A&A...254L...5G} {254, L5}

\bibitem[\protect\citeauthoryear{{Hanuschik}}{{Hanuschik}}{1996}]{Hanuschik96}
{Hanuschik} R.~W.,  1996, \aap, \href
  {https://ui.adsabs.harvard.edu/abs/1996A&A...308..170H} {308, 170}

\bibitem[\protect\citeauthoryear{{Hastings}, {Wang}  \& {Langer}}{{Hastings}
  et~al.}{2020}]{hastings20}
{Hastings} B.,  {Wang} C.,   {Langer} N.,  2020, \mn@doi [\aap]
  {10.1051/0004-6361/201937018}, \href
  {https://ui.adsabs.harvard.edu/abs/2020A&A...633A.165H} {633, A165}

\bibitem[\protect\citeauthoryear{{Hastings}, {Langer}, {Wang}, {Schootemeijer}
  \& {Milone}}{{Hastings} et~al.}{2021}]{hastings21}
{Hastings} B.,  {Langer} N.,  {Wang} C.,  {Schootemeijer} A.,   {Milone} A.~P.,
   2021, \mn@doi [\aap] {10.1051/0004-6361/202141269}, \href
  {https://ui.adsabs.harvard.edu/abs/2021A&A...653A.144H} {653, A144}

\bibitem[\protect\citeauthoryear{{Haubois}, {Carciofi}, {Rivinius}, {Okazaki}
  \& {Bjorkman}}{{Haubois} et~al.}{2012}]{haubois12}
{Haubois} X.,  {Carciofi} A.~C.,  {Rivinius} T.,  {Okazaki} A.~T.,   {Bjorkman}
  J.~E.,  2012, \mn@doi [\apj] {10.1088/0004-637X/756/2/156}, \href
  {https://ui.adsabs.harvard.edu/abs/2012ApJ...756..156H} {756, 156}

\bibitem[\protect\citeauthoryear{{Healy}, {McCullough}  \&
  {Schlaufman}}{{Healy} et~al.}{2021}]{healy21}
{Healy} B.~F.,  {McCullough} P.~R.,   {Schlaufman} K.~C.,  2021, \mn@doi [\apj]
  {10.3847/1538-4357/ac281d}, \href
  {https://ui.adsabs.harvard.edu/abs/2021ApJ...923...23H} {923, 23}

\bibitem[\protect\citeauthoryear{{Hummel}, {Szeifert}, {G{\"a}ssler},
  {Muschielok}, {Seifert}, {Appenzeller}  \& {Rupprecht}}{{Hummel}
  et~al.}{1999}]{hummel99}
{Hummel} W.,  {Szeifert} T.,  {G{\"a}ssler} W.,  {Muschielok} B.,  {Seifert}
  W.,  {Appenzeller} I.,   {Rupprecht} G.,  1999, \aap, \href
  {https://ui.adsabs.harvard.edu/abs/1999A&A...352L..31H} {352, L31}

\bibitem[\protect\citeauthoryear{{Husser} et~al.,}{{Husser}
  et~al.}{2016}]{spexxy}
{Husser} T.-O.,  et~al., 2016, \mn@doi [\aap] {10.1051/0004-6361/201526949},
  \href {https://ui.adsabs.harvard.edu/abs/2016A&A...588A.148H} {588, A148}

\bibitem[\protect\citeauthoryear{{Kamann}, {Wisotzki}  \& {Roth}}{{Kamann}
  et~al.}{2013}]{2013A&A...549A..71K}
{Kamann} S.,  {Wisotzki} L.,   {Roth} M.~M.,  2013, \mn@doi [\aap]
  {10.1051/0004-6361/201220476}, \href
  {https://ui.adsabs.harvard.edu/abs/2013A&A...549A..71K} {549, A71}

\bibitem[\protect\citeauthoryear{{Kamann} et~al.,}{{Kamann}
  et~al.}{2020}]{kamann_ngc1846}
{Kamann} S.,  et~al., 2020, \mn@doi [\mnras] {10.1093/mnras/stz3583}, \href
  {https://ui.adsabs.harvard.edu/abs/2020MNRAS.492.2177K} {492, 2177}

\bibitem[\protect\citeauthoryear{{Kamann}, {Bastian}, {Usher}, {Cabrera-Ziri}
  \& {Saracino}}{{Kamann} et~al.}{2021}]{kamann21}
{Kamann} S.,  {Bastian} N.,  {Usher} C.,  {Cabrera-Ziri} I.,   {Saracino} S.,
  2021, \mn@doi [\mnras] {10.1093/mnras/stab2643}, \href
  {https://ui.adsabs.harvard.edu/abs/2021MNRAS.tmp.2375K} {}

\bibitem[\protect\citeauthoryear{{Kenyon} \& {Hartmann}}{{Kenyon} \&
  {Hartmann}}{1995}]{kenyon95}
{Kenyon} S.~J.,  {Hartmann} L.,  1995, \mn@doi [\apjs] {10.1086/192235}, \href
  {https://ui.adsabs.harvard.edu/abs/1995ApJS..101..117K} {101, 117}

\bibitem[\protect\citeauthoryear{{King}}{{King}}{1962}]{king1962}
{King} I.,  1962, \mn@doi [\aj] {10.1086/108756}, \href
  {https://ui.adsabs.harvard.edu/abs/1962AJ.....67..471K} {67, 471}

\bibitem[\protect\citeauthoryear{{Klement} et~al.,}{{Klement}
  et~al.}{2019}]{klement2019}
{Klement} R.,  et~al., 2019, \mn@doi [\apj] {10.3847/1538-4357/ab48e7}, \href
  {https://ui.adsabs.harvard.edu/abs/2019ApJ...885..147K} {885, 147}

\bibitem[\protect\citeauthoryear{{Klement} et~al.,}{{Klement}
  et~al.}{2022}]{klement22}
{Klement} R.,  et~al., 2022, \mn@doi [\apj] {10.3847/1538-4357/ac4266}, \href
  {https://ui.adsabs.harvard.edu/abs/2022ApJ...926..213K} {926, 213}

\bibitem[\protect\citeauthoryear{{Labadie-Bartz} et~al.,}{{Labadie-Bartz}
  et~al.}{2018}]{labadie18}
{Labadie-Bartz} J.,  et~al., 2018, \mn@doi [\aj] {10.3847/1538-3881/aa9c7e},
  \href {https://ui.adsabs.harvard.edu/abs/2018AJ....155...53L} {155, 53}

\bibitem[\protect\citeauthoryear{{Lee}, {Osaki}  \& {Saio}}{{Lee}
  et~al.}{1991}]{lee1991}
{Lee} U.,  {Osaki} Y.,   {Saio} H.,  1991, \mn@doi [\mnras]
  {10.1093/mnras/250.2.432}, \href
  {https://ui.adsabs.harvard.edu/abs/1991MNRAS.250..432L} {250, 432}

\bibitem[\protect\citeauthoryear{{Lennon}, {Lee}, {Dufton}  \&
  {Ryans}}{{Lennon} et~al.}{2005}]{lennon05}
{Lennon} D.~J.,  {Lee} J.~K.,  {Dufton} P.~L.,   {Ryans} R.~S.~I.,  2005,
  \mn@doi [\aap] {10.1051/0004-6361:20041653}, \href
  {https://ui.adsabs.harvard.edu/abs/2005A&A...438..265L} {438, 265}

\bibitem[\protect\citeauthoryear{{Lim}, {Rauw}, {Naz{\'e}}, {Sung}, {Hwang}  \&
  {Park}}{{Lim} et~al.}{2019}]{lim19}
{Lim} B.,  {Rauw} G.,  {Naz{\'e}} Y.,  {Sung} H.,  {Hwang} N.,   {Park} B.-G.,
  2019, \mn@doi [Nature Astronomy] {10.1038/s41550-018-0619-5}, \href
  {https://ui.adsabs.harvard.edu/abs/2019NatAs...3...76L} {3, 76}

\bibitem[\protect\citeauthoryear{{Lipatov}, {Brandt}  \& {Gossage}}{{Lipatov}
  et~al.}{2022}]{lipatov22}
{Lipatov} M.,  {Brandt} T.~D.,   {Gossage} S.,  2022, \mn@doi [\apj]
  {10.3847/1538-4357/ac78e1}, \href
  {https://ui.adsabs.harvard.edu/abs/2022ApJ...934..105L} {934, 105}

\bibitem[\protect\citeauthoryear{{Mackey} \& {Broby Nielsen}}{{Mackey} \&
  {Broby Nielsen}}{2007}]{2007MNRAS.379..151M}
{Mackey} A.~D.,  {Broby Nielsen} P.,  2007, \mn@doi [\mnras]
  {10.1111/j.1365-2966.2007.11915.x}, \href
  {https://ui.adsabs.harvard.edu/abs/2007MNRAS.379..151M} {379, 151}

\bibitem[\protect\citeauthoryear{{Marino}, {Przybilla}, {Milone}, {Da Costa},
  {D'Antona}, {Dotter}  \& {Dupree}}{{Marino} et~al.}{2018a}]{marino18}
{Marino} A.~F.,  {Przybilla} N.,  {Milone} A.~P.,  {Da Costa} G.,  {D'Antona}
  F.,  {Dotter} A.,   {Dupree} A.,  2018a, \mn@doi [\aj]
  {10.3847/1538-3881/aad3cd}, \href
  {https://ui.adsabs.harvard.edu/abs/2018AJ....156..116M} {156, 116}

\bibitem[\protect\citeauthoryear{{Marino}, {Milone}, {Casagrande}, {Przybilla},
  {Balaguer-N{\'u}{\~n}ez}, {Di Criscienzo}, {Serenelli}  \&
  {Vilardell}}{{Marino} et~al.}{2018b}]{marino18_m11}
{Marino} A.~F.,  {Milone} A.~P.,  {Casagrande} L.,  {Przybilla} N.,
  {Balaguer-N{\'u}{\~n}ez} L.,  {Di Criscienzo} M.,  {Serenelli} A.,
  {Vilardell} F.,  2018b, \mn@doi [\apjl] {10.3847/2041-8213/aad868}, \href
  {https://ui.adsabs.harvard.edu/abs/2018ApJ...863L..33M} {863, L33}

\bibitem[\protect\citeauthoryear{{Martocchia} et~al.,}{{Martocchia}
  et~al.}{2020}]{martocchia2020}
{Martocchia} S.,  et~al., 2020, \mn@doi [\mnras] {10.1093/mnras/staa2929},
  \href {https://ui.adsabs.harvard.edu/abs/2020MNRAS.499.1200M} {499, 1200}

\bibitem[\protect\citeauthoryear{{McSwain} \& {Gies}}{{McSwain} \&
  {Gies}}{2005}]{mcswain05}
{McSwain} M.~V.,  {Gies} D.~R.,  2005, \mn@doi [\apjs] {10.1086/432757}, \href
  {https://ui.adsabs.harvard.edu/abs/2005ApJS..161..118M} {161, 118}

\bibitem[\protect\citeauthoryear{{Milone} et~al.,}{{Milone}
  et~al.}{2017}]{milone_1866}
{Milone} A.~P.,  et~al., 2017, \mn@doi [\mnras] {10.1093/mnras/stw2965}, \href
  {https://ui.adsabs.harvard.edu/abs/2017MNRAS.465.4363M} {465, 4363}

\bibitem[\protect\citeauthoryear{{Milone} et~al.,}{{Milone}
  et~al.}{2018}]{milone18_be}
{Milone} A.~P.,  et~al., 2018, \mn@doi [\mnras] {10.1093/mnras/sty661}, \href
  {https://ui.adsabs.harvard.edu/abs/2018MNRAS.477.2640M} {477, 2640}

\bibitem[\protect\citeauthoryear{{Mucciarelli}, {Dalessandro}, {Ferraro},
  {Origlia}  \& {Lanzoni}}{{Mucciarelli} et~al.}{2014}]{mucciarelli14}
{Mucciarelli} A.,  {Dalessandro} E.,  {Ferraro} F.~R.,  {Origlia} L.,
  {Lanzoni} B.,  2014, \mn@doi [\apjl] {10.1088/2041-8205/793/1/L6}, \href
  {https://ui.adsabs.harvard.edu/abs/2014ApJ...793L...6M} {793, L6}

\bibitem[\protect\citeauthoryear{{Naz{\'e}}, {Rauw}, {Czesla}, {Smith}  \&
  {Robrade}}{{Naz{\'e}} et~al.}{2022}]{naze2022}
{Naz{\'e}} Y.,  {Rauw} G.,  {Czesla} S.,  {Smith} M.~A.,   {Robrade} J.,  2022,
  \mn@doi [\mnras] {10.1093/mnras/stab3378}, \href
  {https://ui.adsabs.harvard.edu/abs/2022MNRAS.510.2286N} {510, 2286}

\bibitem[\protect\citeauthoryear{{Newville}, {Stensitzki}, {Allen}, {Rawlik},
  {Ingargiola}  \& {Nelson}}{{Newville} et~al.}{2016}]{lmfit}
{Newville} M.,  {Stensitzki} T.,  {Allen} D.~B.,  {Rawlik} M.,  {Ingargiola}
  A.,   {Nelson} A.,  2016, {Lmfit: Non-Linear Least-Square Minimization and
  Curve-Fitting for Python}, Astrophysics Source Code Library, record
  ascl:1606.014 (\mn@eprint {ascl} {1606.014})

\bibitem[\protect\citeauthoryear{{Niederhofer}, {Georgy}, {Bastian}  \&
  {Ekstr{\"o}m}}{{Niederhofer} et~al.}{2015}]{niederhofer2015}
{Niederhofer} F.,  {Georgy} C.,  {Bastian} N.,   {Ekstr{\"o}m} S.,  2015,
  \mn@doi [\mnras] {10.1093/mnras/stv1791}, \href
  {https://ui.adsabs.harvard.edu/abs/2015MNRAS.453.2070N} {453, 2070}

\bibitem[\protect\citeauthoryear{{Paxton}, {Bildsten}, {Dotter}, {Herwig},
  {Lesaffre}  \& {Timmes}}{{Paxton} et~al.}{2011}]{mesa}
{Paxton} B.,  {Bildsten} L.,  {Dotter} A.,  {Herwig} F.,  {Lesaffre} P.,
  {Timmes} F.,  2011, \mn@doi [\apjs] {10.1088/0067-0049/192/1/3}, \href
  {https://ui.adsabs.harvard.edu/abs/2011ApJS..192....3P} {192, 3}

\bibitem[\protect\citeauthoryear{{Plummer}}{{Plummer}}{1911}]{plummer1911}
{Plummer} H.~C.,  1911, \mn@doi [\mnras] {10.1093/mnras/71.5.460}, \href
  {https://ui.adsabs.harvard.edu/abs/1911MNRAS..71..460P} {71, 460}

\bibitem[\protect\citeauthoryear{{Pols}, {Cote}, {Waters}  \& {Heise}}{{Pols}
  et~al.}{1991}]{pols1991}
{Pols} O.~R.,  {Cote} J.,  {Waters} L.~B.~F.~M.,   {Heise} J.,  1991, \aap,
  \href {https://ui.adsabs.harvard.edu/abs/1991A&A...241..419P} {241, 419}

\bibitem[\protect\citeauthoryear{{Reig}}{{Reig}}{2011}]{reig2011}
{Reig} P.,  2011, \mn@doi [\apss] {10.1007/s10509-010-0575-8}, \href
  {https://ui.adsabs.harvard.edu/abs/2011Ap&SS.332....1R} {332, 1}

\bibitem[\protect\citeauthoryear{{Rey-Raposo} \& {Read}}{{Rey-Raposo} \&
  {Read}}{2018}]{reyraposo18}
{Rey-Raposo} R.,  {Read} J.~I.,  2018, \mn@doi [\mnras]
  {10.1093/mnrasl/sly150}, \href
  {https://ui.adsabs.harvard.edu/abs/2018MNRAS.481L..16R} {481, L16}

\bibitem[\protect\citeauthoryear{{R{\'\i}mulo} et~al.,}{{R{\'\i}mulo}
  et~al.}{2018}]{rimulo18}
{R{\'\i}mulo} L.~R.,  et~al., 2018, \mn@doi [\mnras] {10.1093/mnras/sty431},
  \href {https://ui.adsabs.harvard.edu/abs/2018MNRAS.476.3555R} {476, 3555}

\bibitem[\protect\citeauthoryear{{Rivinius}, {{\v{S}}tefl}  \&
  {Baade}}{{Rivinius} et~al.}{2006}]{rivinius06}
{Rivinius} T.,  {{\v{S}}tefl} S.,   {Baade} D.,  2006, \mn@doi [\aap]
  {10.1051/0004-6361:20053008}, \href
  {https://ui.adsabs.harvard.edu/abs/2006A&A...459..137R} {459, 137}

\bibitem[\protect\citeauthoryear{{Rivinius}, {Carciofi}  \&
  {Martayan}}{{Rivinius} et~al.}{2013}]{rivinius2013}
{Rivinius} T.,  {Carciofi} A.~C.,   {Martayan} C.,  2013, \mn@doi [\aapr]
  {10.1007/s00159-013-0069-0}, \href
  {https://ui.adsabs.harvard.edu/abs/2013A&ARv..21...69R} {21, 69}

\bibitem[\protect\citeauthoryear{{Saracino} et~al.,}{{Saracino}
  et~al.}{2020}]{saracino2020}
{Saracino} S.,  et~al., 2020, \mn@doi [\mnras] {10.1093/mnras/staa2748}, \href
  {https://ui.adsabs.harvard.edu/abs/2020MNRAS.498.4472S} {498, 4472}

\bibitem[\protect\citeauthoryear{{Saracino} et~al.,}{{Saracino}
  et~al.}{2022}]{saracino2022}
{Saracino} S.,  et~al., 2022, \mn@doi [\mnras] {10.1093/mnras/stab3159}, \href
  {https://ui.adsabs.harvard.edu/abs/2022MNRAS.511.2914S} {511, 2914}

\bibitem[\protect\citeauthoryear{{Shao} \& {Li}}{{Shao} \&
  {Li}}{2014}]{shao2014}
{Shao} Y.,  {Li} X.-D.,  2014, \mn@doi [\apj] {10.1088/0004-637X/796/1/37},
  \href {https://ui.adsabs.harvard.edu/abs/2014ApJ...796...37S} {796, 37}

\bibitem[\protect\citeauthoryear{{Slettebak}}{{Slettebak}}{1982}]{slettebak82}
{Slettebak} A.,  1982, \mn@doi [\apjs] {10.1086/190820}, \href
  {https://ui.adsabs.harvard.edu/abs/1982ApJS...50...55S} {50, 55}

\bibitem[\protect\citeauthoryear{{Sollima}, {D'Orazi}, {Gratton}, {Carini},
  {Carretta}, {Bragaglia}  \& {Lucatello}}{{Sollima} et~al.}{2022}]{sollima22}
{Sollima} A.,  {D'Orazi} V.,  {Gratton} R.,  {Carini} R.,  {Carretta} E.,
  {Bragaglia} A.,   {Lucatello} S.,  2022, \mn@doi [\aap]
  {10.1051/0004-6361/202142928}, \href
  {https://ui.adsabs.harvard.edu/abs/2022A&A...661A..69S} {661, A69}

\bibitem[\protect\citeauthoryear{{Song}, {Mateo}, {Bailey}, {Walker},
  {Roederer}, {Olszewski}, {Reiter}  \& {Kremin}}{{Song}
  et~al.}{2021}]{song2021}
{Song} Y.-Y.,  {Mateo} M.,  {Bailey} John~I. I.,  {Walker} M.~G.,  {Roederer}
  I.~U.,  {Olszewski} E.~W.,  {Reiter} M.,   {Kremin} A.,  2021, \mn@doi
  [\mnras] {10.1093/mnras/stab1065}, \href
  {https://ui.adsabs.harvard.edu/abs/2021MNRAS.504.4160S} {504, 4160}

\bibitem[\protect\citeauthoryear{{Stoehr} et~al.,}{{Stoehr}
  et~al.}{2008}]{stoehr08}
{Stoehr} F.,  et~al., 2008, in {Argyle} R.~W.,  {Bunclark} P.~S.,   {Lewis}
  J.~R.,  eds,  Astronomical Society of the Pacific Conference Series Vol. 394,
  Astronomical Data Analysis Software and Systems XVII. p.~505

\bibitem[\protect\citeauthoryear{{Sun}, {de Grijs}, {Deng}  \& {Albrow}}{{Sun}
  et~al.}{2019}]{sun19}
{Sun} W.,  {de Grijs} R.,  {Deng} L.,   {Albrow} M.~D.,  2019, \mn@doi [\apj]
  {10.3847/1538-4357/ab16e4}, \href
  {https://ui.adsabs.harvard.edu/abs/2019ApJ...876..113S} {876, 113}

\bibitem[\protect\citeauthoryear{{Sun}, {Duan}, {Deng}  \& {de Grijs}}{{Sun}
  et~al.}{2021}]{sun21}
{Sun} W.,  {Duan} X.-W.,  {Deng} L.,   {de Grijs} R.,  2021, \mn@doi [\apj]
  {10.3847/1538-4357/ac1ad0}, \href
  {https://ui.adsabs.harvard.edu/abs/2021ApJ...921..145S} {921, 145}

\bibitem[\protect\citeauthoryear{{Tonry} \& {Davis}}{{Tonry} \&
  {Davis}}{1979}]{tonry79}
{Tonry} J.,  {Davis} M.,  1979, \mn@doi [\aj] {10.1086/112569}, \href
  {https://ui.adsabs.harvard.edu/abs/1979AJ.....84.1511T} {84, 1511}

\bibitem[\protect\citeauthoryear{{Townsend}, {Owocki}  \& {Howarth}}{{Townsend}
  et~al.}{2004}]{townsend04}
{Townsend} R.~H.~D.,  {Owocki} S.~P.,   {Howarth} I.~D.,  2004, \mn@doi
  [\mnras] {10.1111/j.1365-2966.2004.07627.x}, \href
  {https://ui.adsabs.harvard.edu/abs/2004MNRAS.350..189T} {350, 189}

\bibitem[\protect\citeauthoryear{{Wang}, {Langer}, {Schootemeijer}, {Castro},
  {Adscheid}, {Marchant}  \& {Hastings}}{{Wang} et~al.}{2020}]{wang2020}
{Wang} C.,  {Langer} N.,  {Schootemeijer} A.,  {Castro} N.,  {Adscheid} S.,
  {Marchant} P.,   {Hastings} B.,  2020, \mn@doi [\apjl]
  {10.3847/2041-8213/ab6171}, \href
  {https://ui.adsabs.harvard.edu/abs/2020ApJ...888L..12W} {888, L12}

\bibitem[\protect\citeauthoryear{{Wang}, {Gies}, {Peters}, {G{\"o}tberg},
  {Chojnowski}, {Lester}  \& {Howell}}{{Wang} et~al.}{2021}]{wang2021}
{Wang} L.,  {Gies} D.~R.,  {Peters} G.~J.,  {G{\"o}tberg} Y.,  {Chojnowski}
  S.~D.,  {Lester} K.~V.,   {Howell} S.~B.,  2021, \mn@doi [\aj]
  {10.3847/1538-3881/abf144}, \href
  {https://ui.adsabs.harvard.edu/abs/2021AJ....161..248W} {161, 248}

\bibitem[\protect\citeauthoryear{{Wang} et~al.,}{{Wang}
  et~al.}{2022}]{wang2022}
{Wang} C.,  et~al., 2022, \mn@doi [Nature Astronomy]
  {10.1038/s41550-021-01597-5}, \href
  {https://ui.adsabs.harvard.edu/abs/2022NatAs.tmp...35W} {}

\bibitem[\protect\citeauthoryear{{Watkins}, {van de Ven}, {den Brok}  \& {van
  den Bosch}}{{Watkins} et~al.}{2013}]{watkins2013}
{Watkins} L.~L.,  {van de Ven} G.,  {den Brok} M.,   {van den Bosch} R. C.~E.,
  2013, \mn@doi [\mnras] {10.1093/mnras/stt1756}, \href
  {https://ui.adsabs.harvard.edu/abs/2013MNRAS.436.2598W} {436, 2598}

\bibitem[\protect\citeauthoryear{{Weilbacher} et~al.,}{{Weilbacher}
  et~al.}{2020}]{muse_pipeline}
{Weilbacher} P.~M.,  et~al., 2020, \mn@doi [\aap]
  {10.1051/0004-6361/202037855}, \href
  {https://ui.adsabs.harvard.edu/abs/2020A&A...641A..28W} {641, A28}

\bibitem[\protect\citeauthoryear{{Yang}, {Li}, {Deng}, {de Grijs}  \&
  {Milone}}{{Yang} et~al.}{2018}]{yang2018}
{Yang} Y.,  {Li} C.,  {Deng} L.,  {de Grijs} R.,   {Milone} A.~P.,  2018,
  \mn@doi [\apj] {10.3847/1538-4357/aabe26}, \href
  {https://ui.adsabs.harvard.edu/abs/2018ApJ...859...98Y} {859, 98}

\bibitem[\protect\citeauthoryear{{de Mink}, {Langer}, {Izzard}, {Sana}  \& {de
  Koter}}{{de Mink} et~al.}{2013}]{demink2013}
{de Mink} S.~E.,  {Langer} N.,  {Izzard} R.~G.,  {Sana} H.,   {de Koter} A.,
  2013, \mn@doi [\apj] {10.1088/0004-637X/764/2/166}, \href
  {https://ui.adsabs.harvard.edu/abs/2013ApJ...764..166D} {764, 166}

\bibitem[\protect\citeauthoryear{{de Wit}, {Lamers}, {Marquette}  \&
  {Beaulieu}}{{de Wit} et~al.}{2006}]{dewit06}
{de Wit} W.~J.,  {Lamers} H.~J.~G.~L.~M.,  {Marquette} J.~B.,   {Beaulieu}
  J.~P.,  2006, \mn@doi [\aap] {10.1051/0004-6361:20065137}, \href
  {https://ui.adsabs.harvard.edu/abs/2006A&A...456.1027D} {456, 1027}

\makeatother
\end{thebibliography}




\appendix

\section{Calibration of the line broadening provided by Spexxy}
\label{app:sigma2vsini}

In order to convert the Gaussian line broadening provided by \textsc{Spexxy} to \vsini{}, we followed a similar approach as in our previous work on the intermediate-age cluster NGC~1846 \citep{kamann_ngc1846} (see the appendix of that paper). From the second data release of the X-Shooter spectral library \citep[XSL,][]{gonneau2020}, we selected three stars for which \citet{arentsen2019} derived similar stellar parameters to those expected for the upper main sequence of NGC~1850: HD~175640 ($T_{\rm eff}=12438\,{\rm K}$, $\log g=3.99$, ${\rm [Fe/H}]=0.24$), HD~176301 ($T_{\rm eff}=14552\,{\rm K}$, $\log g=3.77$, ${\rm [Fe/H}]=0.26$), and HD~196426 ($T_{\rm eff}=13265\,{\rm K}$, $\log g=3.87$, ${\rm [Fe/H}]=0.23$). Using the XSL data of all three stars as input, we created mock MUSE spectra in the same fashion as described in Sec.~\ref{sec:rotation:combined}. For each mock spectrum, we randomly selected one of the three XSL sources as input, convolved it with a wavelength-dependent model for the MUSE LSF, added a rotational broadening (randomly chosen between 0 and $500\,{\rm km\,s^{-1}}$) using the \textsc{PyAstronomy} \citep{pya} package, and resampled the result to the MUSE wavelength range and spectral sampling. Finally, we added noise by picking a random star from the observed MUSE sample, reading its uncertainty spectrum, and applying it to the mock spectrum. This process was repeated until a final sample of 300 mock MUSE spectra was available. This sample was analysed with \textsc{Spexxy} in the same way as the observed spectra.

\begin{figure}
    \centering
    \includegraphics[width=.95\linewidth]{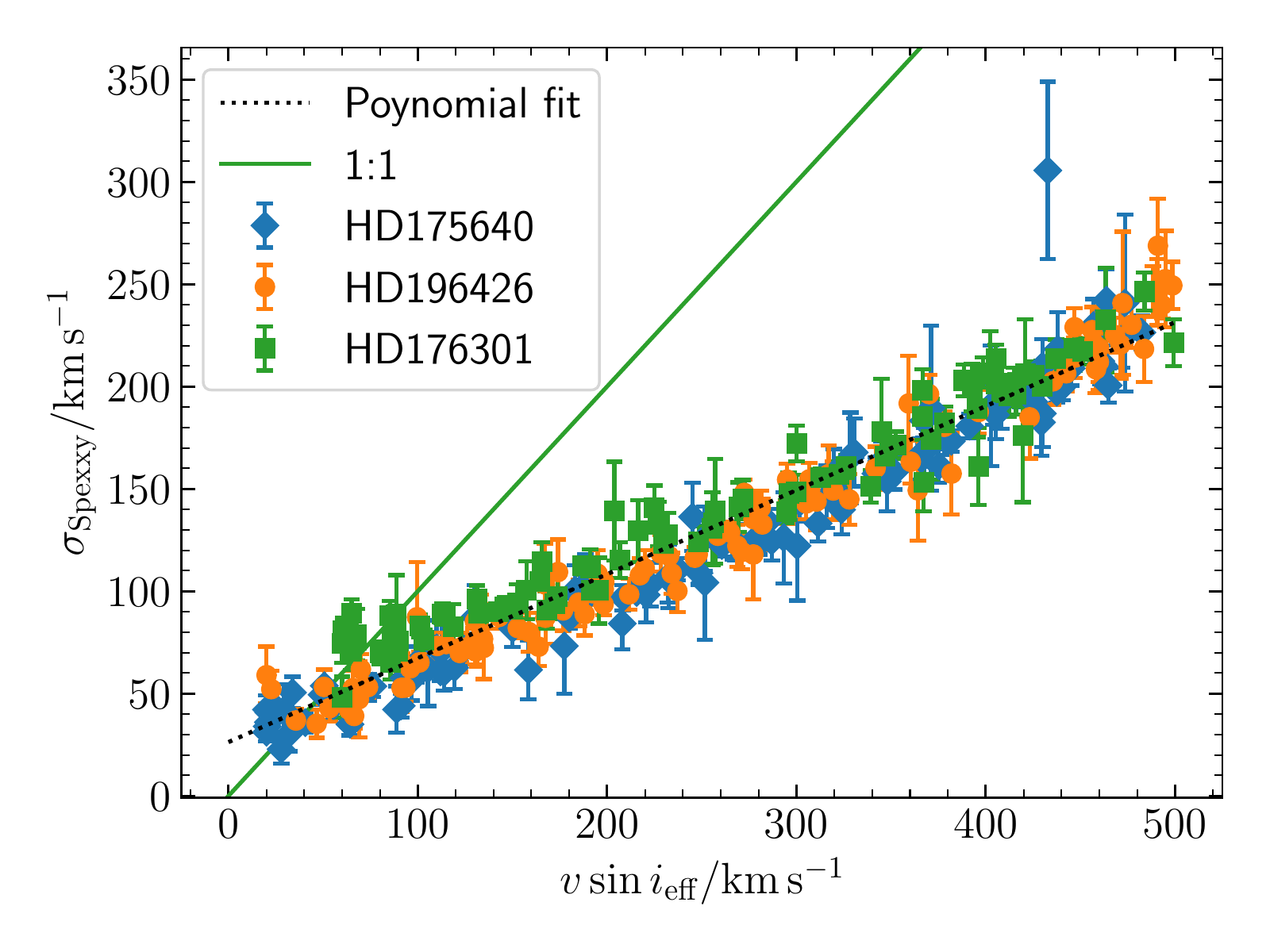}
    \caption{Relation between the effective \vsini{} (i.e. the values obtained by adding the observed rotation rates of the input stars and the ones used in the simulation in quadrature) and the line broadening as determined by \textsc{Spexxy} for 300 mock MUSE spectra. The mock spectra are based on XSL spectra of the three stars named in the legend. The green solid line shows a one-to-one relation while the black dashed line show a first-order polynomial fit to the data.}
    \label{fig:app:vsini}
\end{figure}

When measuring the line broadening, we need to account for the intrinsic rotation of the selected stars. \citet{abt02} measured \vsini{} of the three stars as $20\,{\rm km\,s^{-1}}$ (HD~175640), $60\,{\rm km\,s^{-1}}$ (HD~176301), and $20\,{\rm km\,s^{-1}}$ (HD~196426). We added these values in quadrature to the random \vsini{} values used when creating the mock spectra, resulting in effective \vsini{} values that we compared to the fitted line broadening.\footnote{\citet{bailey13} find a lower value of $\vsini=1.5\,{\rm km\,s^{-1}}$ for HD~175640. However, using this value instead does not significantly alter the relation shown in Fig.~\ref{fig:app:vsini}.}

In Fig.~\ref{fig:app:vsini}, we show the relation between the effective \vsini{} of the mock spectra and the line broadening measured by \textsc{Spexxy} during the analysis. The observed correlation is linear across the entire range of simulated \vsini{} values and no significant differences are observed depending on which star was used to generate a spectrum. Therefore, we fitted the relation with a single polynomial of first order that was then used to convert from the \textsc{Spexxy} output to \vsini{}.

In \citet{kamann_ngc1846} we found that MUSE spectroscopy is insensitive to \vsini{} values $\lesssim 40\,{\rm km\,s^{-1}}$ in early type stars. Accordingly, we set all calculated \vsini{} values below this threshold to $40\,{\rm km\,s^{-1}}$ and consider them as upper limits. This is a slight simplification as the actual threshold will vary from spectrum to spectrum, depending on S/N and stellar type. To get an idea of this variation, we analysed the same simulated spectra described above again with \textsc{Spexxy}, but this time fixed the line broadening to zero. By comparing the reduced $\chi^2$ values between the two fits, we found that for the split main sequence stars, fitting for the line broadening only improves the fits when the simulated \vsini{} is $\gtrsim 70\,{\rm km\,s^{-1}}$. We note that adopting this value as lower threshold instead would not change our conclusions, given that our average \vsini{} for the slow-rotating blue main sequence stars is higher.




\bsp	
\label{lastpage}
\end{document}